\newif\ifconfver
\confvertrue       
\ifconfver
\documentclass[10pt,twocolumn,twoside]{IEEEtran}
\else
\documentclass[11pt,draftcls,onecolumn]{IEEEtran}
\fi

\usepackage{makecell}
\usepackage{bbding}
\usepackage{color,graphicx}
\usepackage{float}
\usepackage{multirow,amsmath,epsfig,amsfonts,amssymb,psfig,graphics,psfrag,theorem,calc,url,bm,cite}
\usepackage{stfloats,hyperref}
\usepackage{algorithm,algorithmic}
\usepackage{diagbox} 
\usepackage{hhline}
\usepackage{subfigure}
\usepackage{physics}
\usepackage{booktabs}
\usepackage{tikz}
\usepackage{multirow}
\usepackage{arydshln}
\usepackage{booktabs, multirow}    
\usepackage{placeins}
\usepackage{graphicx}
\usepackage{subcaption}

\usepackage{mathtools}

\makeatletter
\def\multilimits@{\bgroup
	\Let@
	\restore@math@cr
	\default@tag
	\baselineskip\fontdimen10 \scriptfont\tw@
	\advance\baselineskip\fontdimen12 \scriptfont\tw@
	\lineskip\thr@@\fontdimen8 \scriptfont\thr@@
	\lineskiplimit\lineskip
	\vbox\bgroup\ialign\bgroup\hfil$\m@th\scriptstyle{##}$\hfil\crcr}
\def\Sb{_\multilimits@}
\def\endSb{\crcr\egroup\egroup\egroup}
\makeatother

	{\end{list}}

\newlength{\twidth}
\ifconfver
\else
\fi

\definecolor{orange}{RGB}{255,107,0}

\theorembodyfont{\rmfamily}

{\begin{list}{}{
			\settowidth{\labelwidth}{\mbox{\textnormal{#1}}}%
			\setlength{\leftmargin}{\labelwidth+\labelsep}}}%
	{\end{list}}

\newcommand\bB{\ensuremath{{\bm B}}}

\newcommand\bD{\ensuremath{{\bm D}}}

\newcommand\bH{\ensuremath{{\bm H}}}
\newcommand\bI{\ensuremath{{\bm I}}}

\newcommand\bS{\ensuremath{{\bm S}}}

\newcommand\bd{\ensuremath{{\bm d}}}

\newcommand\bv{\ensuremath{{\bm v}}}

\newcommand\bx{\ensuremath{{\bm x}}}
\newcommand\by{\ensuremath{{\bm y}}}
\newcommand\bz{\ensuremath{{\bm z}}}

\definecolor{orange}{RGB}{255,107,0}

\ifconfver

\author{Chia-Hsiang Lin,~\IEEEmembership{Senior Member,~IEEE}, Wei-Chih Liu,~\IEEEmembership{Student Member,~IEEE},\\ Yu-En Chiu,~\IEEEmembership{Student Member,~IEEE}, and Jhao-Ting Lin,~\IEEEmembership{Student Member,~IEEE}}

\title{Deep Unfolding Real-Time Super-Resolution Using Subpixel-Shift Twin Image and Convex Self-Similarity Prior

\thanks{This study was supported by the Emerging Young Scholar Program (namely, the 2030 Cross-Generation Young Scholars Program) of National Science and Technology Council (NSTC), Taiwan, under Grant NSTC 114-2628-E-006- 002.
We thank the National Center for Theoretical Sciences (NCTS) and the National Center for High-performance Computing (NCHC) for providing the computing resources.
\textit{(Corresponding Author: Chia-Hsiang Lin)}}
\thanks{C.-H. Lin is with the Department of Electrical Engineering, National Cheng Kung University, Tainan 70101, Taiwan
(e-mail: chiahsiang.steven.lin@gmail.com).}
\thanks{W.-C. Liu is with the Institute of Computer and Communication Engineering, Department of Electrical Engineering, National Cheng Kung University, Tainan 70101, Taiwan
(e-mail:  q36131087@gs.ncku.edu.tw).}
\thanks{Y.-E. Chiu is with the Institute of Computer and Communication Engineering, Department of Electrical Engineering, National Cheng Kung University, Tainan 70101, Taiwan
(e-mail:  q36131134@gs.ncku.edu.tw).}
\thanks{J.-T. Lin is with the Institute of Computer and Communication Engineering, Department of Electrical Engineering, National Cheng Kung University, Tainan 70101, Taiwan
(e-mail:  q38091534@gs.ncku.edu.tw).}
}
\else

\fi

\begin{document}
	
	\bibliographystyle{IEEEtran}
	\maketitle
	\ifconfver \else \vspace{-0.5cm}\fi

\begin{abstract}
Multi-image super-resolution (MISR) is a critical technique for satellite remote sensing.
In the perspective of information, twin-image super-resolution (TISR) is regarded as the most challenging MISR scenario, having crucial applications like the SPOT-5 supermode imaging.
In TISR, an image is super-resolved by its subpixel-shift counterpart (i.e., twin image), where the two images are typically offset by half a pixel both horizontally and vertically.
We formulate the less investigated TISR using a convex criterion, which is implemented using a novel deep unfolding network.
In the unfolding, an embedded simple shift operator trickily addresses the coupled TISR data-fitting terms, and a transformer trained with a convex self-similarity loss function elegantly implements the proximal mapping induced by the TISR regularizer.
The proposed convex self-similarity unfolding supermode super-resolution (COSUP) algorithm is interpretable and achieves state-of-the-art performance with very fast millisecond-level computational time.
COSUP is also tested on real-world data, for which the subpixel shifts would not be spatially uniform, with results showing great superiority over the official CNES supermode imaging product in terms of credible metrics (e.g., natural image quality evaluator, NIQE).
Source codes: https://github.com/IHCLab/COSUP.
\end{abstract}

	
	\ifconfver \else \vspace{-0.0cm}\fi
	
	\ifconfver \else \vspace{-0.5cm}\fi
	
	\ifconfver \else  \fi

\section{Introduction}\label{sec: introduction}

Super-resolution (SR) becomes an essential image reconstruction technique in the remote sensing area, and has widespread applications in satellite remote sensing, including urban planning, defense surveillance, precision agriculture, and disaster response.
Generally, to tackle hardware limitations, a common strategy involves fusing multi-modal data, such as fusing low-resolution (LR) hyperspectral images (HSIs) with high-resolution (HR) multispectral images (MSIs) to obtain the desired HR HSIs.
Specifically, advanced deep learning architectures have been widely explored in this domain \cite{CODEIF, cao2024unsupervised, CAO2026112374}.
For instance, a hybrid framework integrating convolutional neural network (CNN) and transformer has been applied to achieve unsupervised HSI-MSI fusion with a degradation-aware design \cite{cao2024unsupervised}.
In the very recent work \cite{CAO2026112374}, a cross-domain-aware transformer is proposed to leverage the spatio-spectral attention prior in the training stage.
In parallel to the multi-modal fusion approaches, multi-image super-resolution (MISR) effectively reconstructs the spatial details \cite{an2022tr,sun2025super} without expensive hardware costs.
%
A representative MISR work proposed in \cite{an2022tr} adopts an end-to-end framework to integrate a residual encoder, a transformer fusion module, and a subpixel decoder.
The transformer dynamically attends to corresponding regions across images, effectively capturing fine details for MISR.
The transformer is trained by ignoring spatial patch relations and shows strong performance on the PROBA-V Kelvin dataset \cite{an2022tr}. 
In the very recent work \cite{sun2025super}, a novel gradient-guided MISR framework for remote sensing imagery is proposed, by combining a learned local gradient prior with a model-based nonlocal total variation (TV) prior using the maximum a posteriori (MAP) strategy. 
These priors complement each other by enhancing fine structures and preserving edge smoothness. 
The overall optimization uses adaptive norm-based reconstruction to solve a mixed L1-L2 minimization problem \cite{sun2025super}.

\begin{figure}[t]
\centering
\hspace*{2.0em} 
\includegraphics[width=0.62\linewidth]{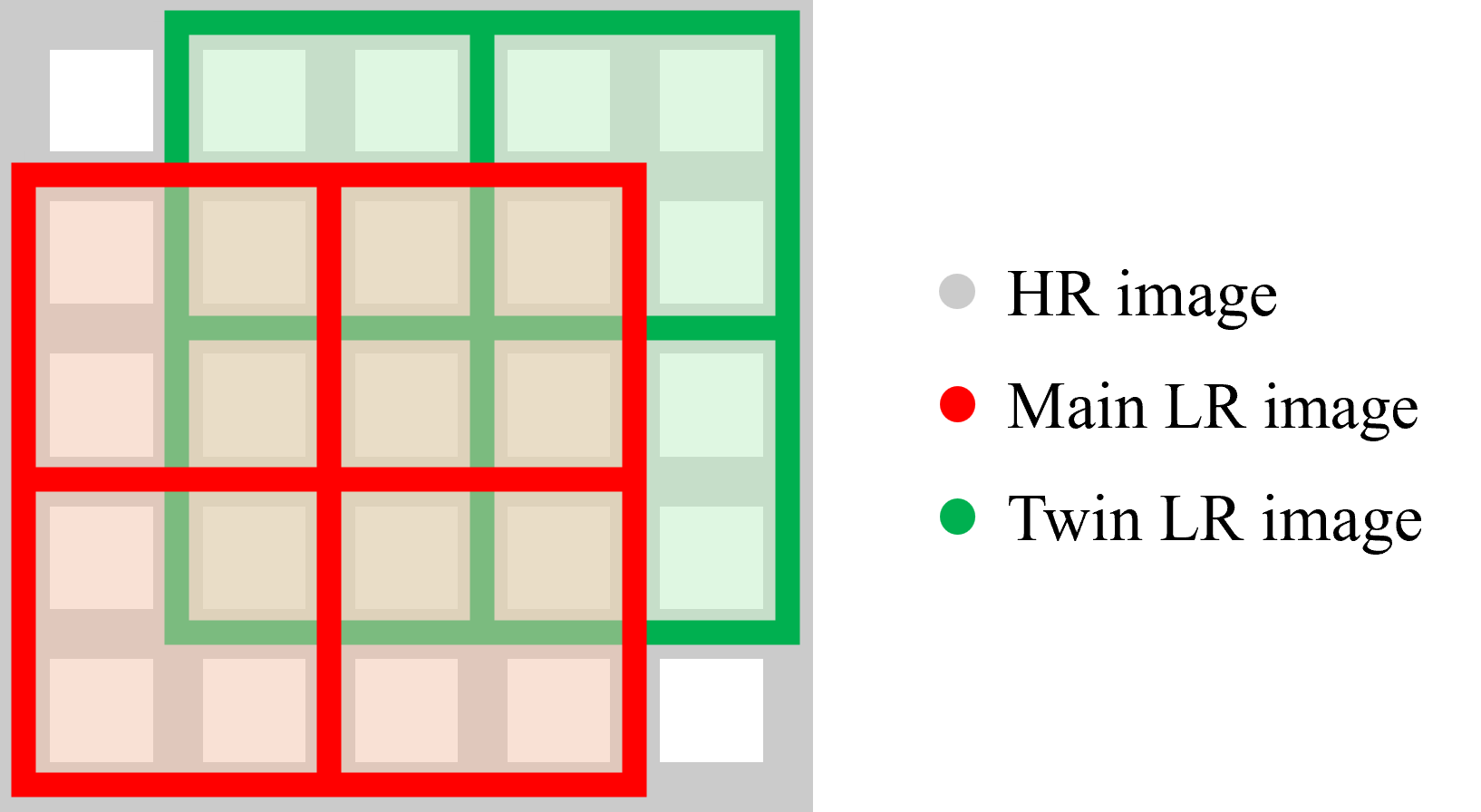}
\caption{Graphical illustration of the considered twin-image super-resolution (TISR) problem.
The unknown target image $\bz$ is the HR image (gray one).
The aim is to super-resolve the LR image $\by_1$ (red one) to obtain the target HR image $\bz$, where the ground sampling distance (GSD) of the LR image is twice that of the target image.
This TISR will be achieved through the help of the subpixel-shift twin image $\by_2$ (green one).
The green and red images are typically offset by half a pixel both horizontally and vertically.}
\label{fig:TISRillustration}
\end{figure}

However, most MISR methods focus on scenarios of SR from more than two images, including 
the MISR from four images with upscaling factor of 2 \cite{dong2019selection,wu2024hybrid}, 
the MISR from nine images with upscaling factor of 3 \cite{molini2019deepsum}, and
the MISR from sixteen images with upscaling factor of 4 \cite{hardie1997joint}.
Compared to the above scenarios, a clearly more challenging case is the MISR from only two images with upscaling factor of 2---the so-called twin-image super-resolution (TISR) problem, as illustrated in Figure \ref{fig:TISRillustration}.

In the perspective of information, TISR is regarded as the most challenging MISR scenario, and has crucial satellite remote sensing applications, such as the SPOT-5 satellite supermode imaging.
In TISR, an LR image (i.e., the red image in Figure \ref{fig:TISRillustration}) is super-resolved by its LR subpixel-shift counterpart, referred to as the twin image (i.e., the green image in Figure \ref{fig:TISRillustration}), to obtain an HR target image (i.e., the gray image in Figure \ref{fig:TISRillustration}).
In the supermode scenario of Centre National d'Études Spatiales (CNES) \cite{latry1998spot5}, the two LR images are typically offset by half a pixel both horizontally and vertically, as illustrated in Figure \ref{fig:TISRillustration}.
There are basically three ways to achieve TISR, including the adoption of the CNES product, the simplification of MISR methods, and the single-image super-resolution (SISR) directly from the LR image, as will be introduced next.

The first and probably the most straightforward one is to use the official CNES SPOT World Heritage (SWH) Level 2A-Carto supermode image product \cite{latry1998spot5}.
As part of the SPOT-5 development program, CNES initiated a study to investigate cost-effective approaches for enhancing the spatial sampling rate of the panchromatic (PAN) 5-meter mode, ultimately resulting in the implementation of the THR (Très Haute Résolution, or Very High Resolution) operational mode.
Although directly using CNES product is convenient, it supports only the SPOT-5 satellite image SR; for images acquired by other satellites (e.g., FORMOSAT-8 satellite \cite{lin2024synthesis}), the CNES platform does not accept the LR image (and its twin image) for further TISR procedure.
In this paper, we aim to propose a generally applicable TISR tool for arbitrary satellite imagery in the form specified by Figure \ref{fig:TISRillustration}.
Our method outperforms the CNES product in the SPOT-5 satellite TISR task on one hand (cf. Section \ref{sec:experiment}), and is generally applicable to other satellite TISR tasks on the other hand (cf. Section \ref{sec:experiment}).

Another TISR approach is to simplify some existing MISR methods, so that they can accept two input images (i.e., the red and green LR images in Figure \ref{fig:TISRillustration}).
As far as we could track, the earliest work in this line \cite{stark1989high} addresses the reconstruction of remotely sensed images from image-plane detector arrays, where detector elements may exceed the size of the optical blur spot. 
By leveraging image shifts through scanning or rotation, MISR become feasible. 
Instead of relying on matrix inversion or least-squares methods, a convex projections approach is proposed, motivating us to adopt the convex self-similarity prior originally proposed in \cite{SSSS}.
The method \cite{stark1989high} effectively achieves high-quality reconstructions with an arbitrary number of input images.
Another representative work \cite{kawulok2019deep} adopts deep learning to achieve MISR.
The work proposes a novel deep evolutionary network (EvoNet) \cite{kawulok2019deep} to fuse multiple images and to learn the LR-to-HR mapping, thereby achieving superior MISR performances.
MISR and image subpixel registration can also be performed simultaneously even without the aid of the target HR reference images \cite{kim2025self}.
MISR with an arbitrary number of LR images (e.g., TISR) has been achieved through image registration \cite{irani1991improving} and game theory-driven attention-based deep learning \cite{wang2023multi}.

The TISR problem can also be solved by directly applying some SISR methods to super-resolve the LR images (e.g., the red image in Figure \ref{fig:TISRillustration}).
Recent advances in CNN have significantly improved remote sensing image SR.
Specifically, \cite{10764782} proposes a multiscale feature fusion network by integrating CNN and transformer to extract local details and global contextual information.
%
In a different kind of approach, \cite{rs15040882} and \cite{9151194} adopt closed-loop network and cycle-CNN, respectively, to fully exploit the LR/HR image textures under the challenging SISR framework.
The work \cite{lei2021hybrid} leverages the internal recurrence of ground targets to enhance feature representation. 
A novel hybrid-scale self-similarity exploitation network is proposed for SISR, incorporating a single-scale self-similarity module and a cross-scale connection structure, thereby capturing both intra- and inter-scale similarities \cite{lei2021hybrid}. 
The idea is related to the multi-scale feature fusion, which has been applied to SISR through quantum deep network (QUEEN) in recent literature \cite{hsu2024hyperqueen}.
QUEEN was originally implemented for quantum reconstruction of HSI in the remote sensing area \cite{HyperQUEEN}, and very recently the technique has been extended to the quantum generative artificial intelligence (AI) using game theory for quantum image reconstruction and noise removal \cite{HyperKING}.
As indicated in \cite{kawulok2019deep}, while deep learning has significantly advanced SISR, its potential remains underexplored in MISR or TISR, which fuses complementary information across images for improved accuracy.

SR deep networks are, however, mostly built based on empirical testing.
In this work, we aim to build interpretable deep network through deep unfolding \cite{COS2A}.
In many deep unfolding networks, researchers often formulate their problems as a convex optimization problems, derive closed-form solutions for the problems, and implement the deep architectures based on the structures of the solutions.
Another completely different way to integrate convex optimization and deep learning for solving ill-posed inverse problems are referred to as the convex/deep (CODE) learning theory \cite{NCCODE}, which was originally invented for satellite image reconstruction as well \cite{CODE}.
CODE theory excels in small-data learning, and has led to outstanding SISR results \cite{SISRlin2022single}.
As mentioned, TISR is fundamentally a fusion technique \cite{COCNMF}, for which CODE theory also achieves image SR via information fusion \cite{CODEIF}.
In this paper, we aim to formulate the ill-posed TISR as a convex problem regularized by image self-similarity prior \cite{buades2011self,burger2012image}.

Specifically, our TISR criterion consists of two convex data-fitting terms (corresponding to the two input LR images) and one convex self-similarity regularizer \cite{SSSS}.
The self-similarity phenomenon was explicitly defined as a convex function in \cite{SSSS} for multi-resolution analysis of the Sentinel-2 satellite data.
Then, we solve the proposed TISR criterion using alternating direction method of multipliers (ADMM) \cite{CVXbookCLL2016}, and implement it using a novel deep unfolding network.
In the unfolding architecture, an embedded simple shift operator trickily addresses the coupled TISR data-fitting terms (cf. Figure \ref{fig:TISRillustration}).
The self-similarity regularization induces a proximal mapping, which is elegantly unfolded as a transformer trained using the convex self-similarity loss function.
For the first time, the convex self-similarity phenomenon is incorporated into the deep learning framework.
The proposed convex self-similarity unfolding supermode super-resolution (COSUP) algorithm is interpretable and computationally efficient, and achieves superior TISR performance over established baselines.
The interpretable COSUP method is also tested on real-world data, for which the subpixel shifts would not be as perfectly aligned as depicted in Figure \ref{fig:TISRillustration}, and the results even outperform the official CNES supermode imaging product in terms of natural image quality evaluator (NIQE).
In summary, the novelty and main contributions are itemized below.
\begin{enumerate}
\item 
Typical MISR approaches rely on more than two images, while the proposed COSUP algorithm requires only two images---the so-called TISR problem (regarded as the most challenging MISR scenario).
For the first time, we formulate the TISR problem as a convex optimization (CO) problem \cite{CVXbookCLL2016,AAHCSD}.
Given the unique challenges and significance of the TISR problem, we solve it via a customized ADMM-based deep unfolding network, where a shift operator is trickily embedded to couple two data-fitting terms, thereby facilitating effective information fusion between two images.
%
%
This is also the first time that a convex self-similarity function is adopted to explicitly regularize the ill-posed TISR problem.
The above theoretical settings allow the adoption of existing powerful CO theory (e.g., ADMM \cite{CVXbookCLL2016}) to elegantly and effectively solve the target problem.

\smallskip


\smallskip

\item 
Unlike the official CNES product, exclusively designed for SPOT-5 images, the proposed COSUP algorithm is a generalized framework applicable to arbitrary satellite imageries (e.g., FORMOSAT-8 satellite \cite{formosat-8}).
TISR problem involves a highly coupled twin-image pair with subpixel shifts, and COSUP is explicitly designed to exploit this structural characteristic, enabling stable reconstruction under both ideal half-pixel shifts and nonideal misregistration conditions, achieving superior performance on simulation data (cf. Section \ref{sec:simulation data}) and real-world data (cf. Section \ref{sec:real data (SPOT-5)}).
We also propose a customized experimental protocol for investigating the nonideal misregistration conditions (cf. Section \ref{sec:ablation}).

\smallskip
 
\item 
To have HR self-similarity pattern, we propose a judicious strategy that directly integrates the self-similarity function (in convex optimization) into the loss function (in deep learning).
Accordingly, we develop a transformer trained via the convex self-similarity loss function to elegantly implement the TISR regularizer. 
This is the first time that the self-similarity phenomenon is explicitly incorporated into a convex/deep learning framework \cite{CODE,COS2A}.
This design elegantly synergizes the intrinsic attention mechanism of the transformer with the mathematics-driven self-similarity prior, thereby enforcing a strict alignment between the learned attention patterns and image self-similarity structures.
This suggests that transformer is inherently an ideal counterpart of our convex self-similarity loss function, leading to state-of-the-art TISR performances.
This concise deployment also leads to the millisecond-level real-time computing of COSUP.
\end{enumerate}

The remaining parts of the paper are organized as below.
In Section \ref{sec:theory}, we formulate TISR as a convex problem, and implement it using deep unfolding, thereby leading to the COSUP algorithm.
In Section \ref{sec:experiment}, we conduct quantitative and qualitative performance evaluations to demonstrate the superiority of the proposed COSUP method.
Concluding remarks are then drawn in Section \ref{sec:conclusion}.

\smallskip 

\textit{Notation:}
$\mathbb{R}^{M}$ and $\mathbb{R}^{M \times N}$ denote the $M$-dimensional Euclidean space and the $(M\times N)$-dimensional matrix space, respectively.
$\| \cdot \|_2$ and $\| \cdot \|_1$ are the Euclidean L2 norm and the L1 norm, respectively.
$|\mathcal{S}|$ denotes the set cardinality of a given set $\mathcal{S}$.
$\bI$ and $\bm 0$ are the identity matrix and the zero vector, respectively.

\section{The Proposed COSUP Algorithm}\label{sec:theory}

\subsection{TISR Problem Formulation}\label{sec:probdesc}

Let $\bz \in \mathbb{R}^{MN}$ represent the target $M\times N$ HR PAN image (i.e., the gray image in Figure \ref{fig:TISRillustration}), where $M$ and $N$ are assumed to be even numbers without loss of generality.
The aim is to estimate $\bz$ from the $m\times n$ LR image $\by_1\in\mathbb{R}^{mn}$ (i.e., the red image in Figure \ref{fig:TISRillustration}), which is acquired over the same scene as $\bz$ with $m = M/2$ and $n = N/2$.
In TISR, the above SR task is achieved through the help of the twin image $\by_2\in\mathbb{R}^{mn}$, which is a subpixel-shift version of $\by_1$ as illustrated by the green image in Figure \ref{fig:TISRillustration}.

According to the above problem description, the relation between $\bz$ and $\by_1$ can be modeled as $\by_1 =\bD\bB\bz$ by following typical SR mathematical modeling \cite{paris2018novel}, in which $\bB \in \mathbb{R}^{MN \times MN}$ and $\bD \in \mathbb{R}^{mn \times MN}$ are the blurring matrix and the downsampling matrix, respectively.
Also, the subpixel-shift twin version can be modeled as $\by_2 = \bD\bB \bS \bz$ with the shifting matrix $\bS \in \mathbb{R}^{MN \times MN}$, meaning that the target image is shifted by an HR pixel (i.e., half an LR pixel) both horizontally and vertically before the blurring/downsampling operations (cf. Figure \ref{fig:TISRillustration}).
Therefore, the TISR problem can be naturally formulated as
\begin{equation} 
\label{eq:regularized_problem}
\min_{\bz} \ \frac{1}{2} \| \bD\bB \bz - \by_1 \|_2^2 + \frac{1}{2} \|\bD \bB \bS \bz - \by_2 \|_2^2 + \lambda f(\bz),
\end{equation}
where $\lambda > 0$ controls the regularization strength, and $f(\cdot)$ denotes a regularization function for mitigating the ill-posed nature of the TISR task.

In this work, we use the self-similarity prior of natural images \cite{xu2015patch} to design the regularization term $f(\cdot)$.
The self-similarity structure has been explicitly defined as a convex function in \cite{SSSS}.
Although there are more frequently seen priors, such as sparsity prior and TV prior, they are not differentiable and scene-adaptable.
Therefore, we design a scene-adapted self-similarity function, which is more general than the definition proposed in \cite{SSSS} (cf. Appendix \ref{appdx:Kalpha}), and our definition is still differentiable and convex.
Specifically, the regularizer $f(\bz)$ in \eqref{eq:regularized_problem} is defined as
\begin{equation}
f(\bz) \triangleq \frac{1}{2} \sum_{(i,j) \in \mathcal{K}} \alpha_{i,j} \left\| \mathbf{P}_i \bz - \mathbf{P}_j \bz \right\|_2^2,
\label{eq:ss_phi}
\end{equation}
where $\mathbf{P}_k \in \mathbb{R}^{q^2 \times MN}$ denotes the $k$th patch-sifting operator that extracts the $k$th $q\times q$ patch from the HR image $\bz$ \cite{SSSS}.
The self-similarity pattern here is described using a more general weighted graph $\mathcal{K}$ with the edge weights given by $\{\alpha_{i,j}\}$ (cf. Appendix \ref{appdx:Kalpha}), where the edge $(i,j)$ is collected into the graph $\mathcal{K}$ if the $i$th and $j$th patches are similar, and the weight of the edge $(i,j)$ is given by the similarity degree of $\alpha_{i,j}\geq 0$.

Unlike the scenario considered in \cite{SSSS}, we do not have an HR reference image in the TISR problem setting for computing the self-similarity pattern $(\mathcal{K},\{\alpha_{i,j}\})$, making the convex regularizer \eqref{eq:ss_phi} not directly applicable to TISR.
However, assuming that the self-similarity pattern $(\mathcal{K},\{\alpha_{i,j}\})$ is known, the physical meaning of \eqref{eq:ss_phi} is to promote the similarity between the $i$th and $j$th patches (i.e., $\mathbf{P}_i\bz$ and $\mathbf{P}_j\bz$) in the final solution $\bz$.
Also, for larger similarity level $\alpha_{i,j}$, such a promotion will also get stronger, as can be seen from \eqref{eq:ss_phi}.
As for how to learn the self-similarity pattern $(\mathcal{K},\{\alpha_{i,j}\})$ in the TISR case, the mathematical details are relegated to Appendix \ref{appdx:Kalpha}.
Echoing the CODE theory \cite{CODE,NCCODE}, in our TISR algorithm development, the convex self-similarity prior \eqref{eq:ss_phi} will be incorporated into deep learning for the first time.

\subsection{The COSUP Algorithm for Solving \eqref{eq:regularized_problem}}\label{sec:algodesign}

To solve \eqref{eq:regularized_problem}, we develop the COSUP algorithm based on the ADMM optimization theory \cite{CVXbookCLL2016}.
To this end, we need reformulate \eqref{eq:regularized_problem} into the ADMM form, which involves a two-term variable-independent objective function with a constraint that linearly associates the two block variables.
Accordingly, we first merge the two data-fitting terms of \eqref{eq:regularized_problem} into a single objective term, i.e.,
\begin{equation}
\|\bH\bz-\by\|_2^2
=
\| \bD\bB \bz - \by_1 \|_2^2 
+ \|\bD \bB \bS \bz - \by_2 \|_2^2,
\end{equation}
where 
$\bH \triangleq \begin{bmatrix} \bD\bB \\ \bD\bB \bS \end{bmatrix} \in \mathbb{R}^{2mn \times MN}$ 
and 
$\by \triangleq \begin{bmatrix} \by_1 \\ \by_2 \end{bmatrix} \in \mathbb{R}^{2mn}$, leading to the two-
term reformulation of \eqref{eq:regularized_problem}, i.e.,
\begin{equation} 
\label{prob:mid}
\min_{\bz} \ \frac{1}{2} \|\bH\bz-\by\|_2^2 + \lambda f(\bz).
\end{equation}
To further transform \eqref{prob:mid} into the standard ADMM form, we introduce a linear variable constraint ``$\bx=\bz$'' to convert the objective function into a  variable-independent one, i.e.,
\begin{equation} 
\label{eq:constrained_problem}
\begin{aligned}
\min_{\bz} \quad & \frac{1}{2} \| \bH \bx - \by \|_2^2 + \lambda f(\bz) \\
\text{s.t.} \quad & \bx = \bz
\end{aligned}
\end{equation}
which can now be solved using the ADMM theory.

Specifically, \eqref{eq:constrained_problem} can be solved based on its augmented Lagrangian \cite{CVXbookCLL2016}, which is defined as
\begin{equation} \label{eq:augmented_lagrangian}
\mathcal{L}(\bx, \bz, \bd) = \frac{1}{2} \| \bH \bx - \by \|_2^2 + \lambda f(\bz) + \frac{c}{2} \| \bx - \bz - \bd \|_2^2,
\end{equation}
where $\bd \in \mathbb{R}^{MN}$ is the scaled dual variable, and $ c > 0$ denotes the penalty parameter in ADMM \cite{CVXbookCLL2016}.
Then, the ADMM-based COSUP algorithm solves \eqref{eq:regularized_problem}, or equivalently \eqref{eq:constrained_problem}, by iteratively updating the following three steps, i.e.,
\begin{align}
\bz^{k+1}
&:=
\arg\min_{\bz}
\mathcal{L}(\bx^k,\bz,\bd^k),
\label{eq:primal_update1}
\\
\bx^{k+1}
&:=
\arg\min_{\bx}
\mathcal{L}(\bx,\bz^{k+1},\bd^k),
\label{eq:primal_update2}
\\
\bd^{k+1} 
&:= 
\bd^k - \left( \bx^{k+1} - \bz^{k+1} \right),
\label{eq:dual_update}
\end{align}
whose closed-form expressions can be derived as follows.
As we need $(\bx^0,\bd^0)$ to initialize the algorithm, we simply set $\bx^0:=\frac{1}{2}\bH^\top\by$ and $\bd^0:=\bm 0_{MN}$, as illustrated in Figure \ref{fig:stage_k}.

First, according to \eqref{eq:augmented_lagrangian} and \eqref{eq:primal_update1}, the $\bz$-update can be explicitly written as
\begin{equation} \label{eq:z_update_problem}
\begin{aligned}
\bz^{k+1}
&= \arg\min_{\bz}\; \lambda f(\bz) + \frac{c}{2}\,\|\bx^k - \bz - \bd^k\|_2^2 
\\
&= \operatorname{prox}_{\frac{\lambda}{c} f}\!\big(\bx^{k}-\bd^{k}\big),
\end{aligned}
\end{equation}
where $\textrm{prox}_{g}(\bv)\triangleq\arg\min_{\bx} g(\bx)+\frac{1}{2}\|\bx-\bv\|_2^2$ denotes the proximal operator associated with a given function $g$ \cite{parikh2014proximal}.
Next, according to \eqref{eq:augmented_lagrangian} and \eqref{eq:primal_update2}, the $\bx$-update can be explicitly written as
\begin{equation} 
\label{eq:x_update_problem}
\bx^{k+1} = \arg \min_{\bx} \ \frac{1}{2} \| \bH \bx - \by \|_2^2 + \frac{c}{2} \| \bx - \bz^{k+1} - \bd^k \|_2^2.
\end{equation}
Setting the gradient of \eqref{eq:x_update_problem} to zero yields a closed-form solution, i.e.,
\begin{equation} \label{eq:x_update_closed}
\bx^{k+1} = \left( \bH^\top \bH + c \bI \right)^{-1} \left( \bH^\top \by + c (\bz^{k+1} + \bd^k) \right).
\end{equation}
To summarize, the proposed COSUP algorithm is to iteratively implement the two primal updates (i.e., \eqref{eq:z_update_problem} and \eqref{eq:x_update_closed}) and one dual update (i.e., \eqref{eq:dual_update}).
The three updates of COSUP algorithm can be implemented using interpretable deep network, as will be detailed later.

\subsection{Deep Unfolding Implementation of the COSUP Algorithm}\label{sec:implementation}

Echoing the convex/deep (CODE) theory \cite{CODE,NCCODE}, the convex TISR algorithm, i.e., COSUP, will be implemented using deep unfolding.
In this subsection, we design lightweight deep architectures corresponding to the three ADMM closed-form solutions \eqref{eq:z_update_problem}, \eqref{eq:x_update_closed}, and \eqref{eq:dual_update}.

\begin{figure}[t]
\centering
\includegraphics[width=1.0\linewidth]{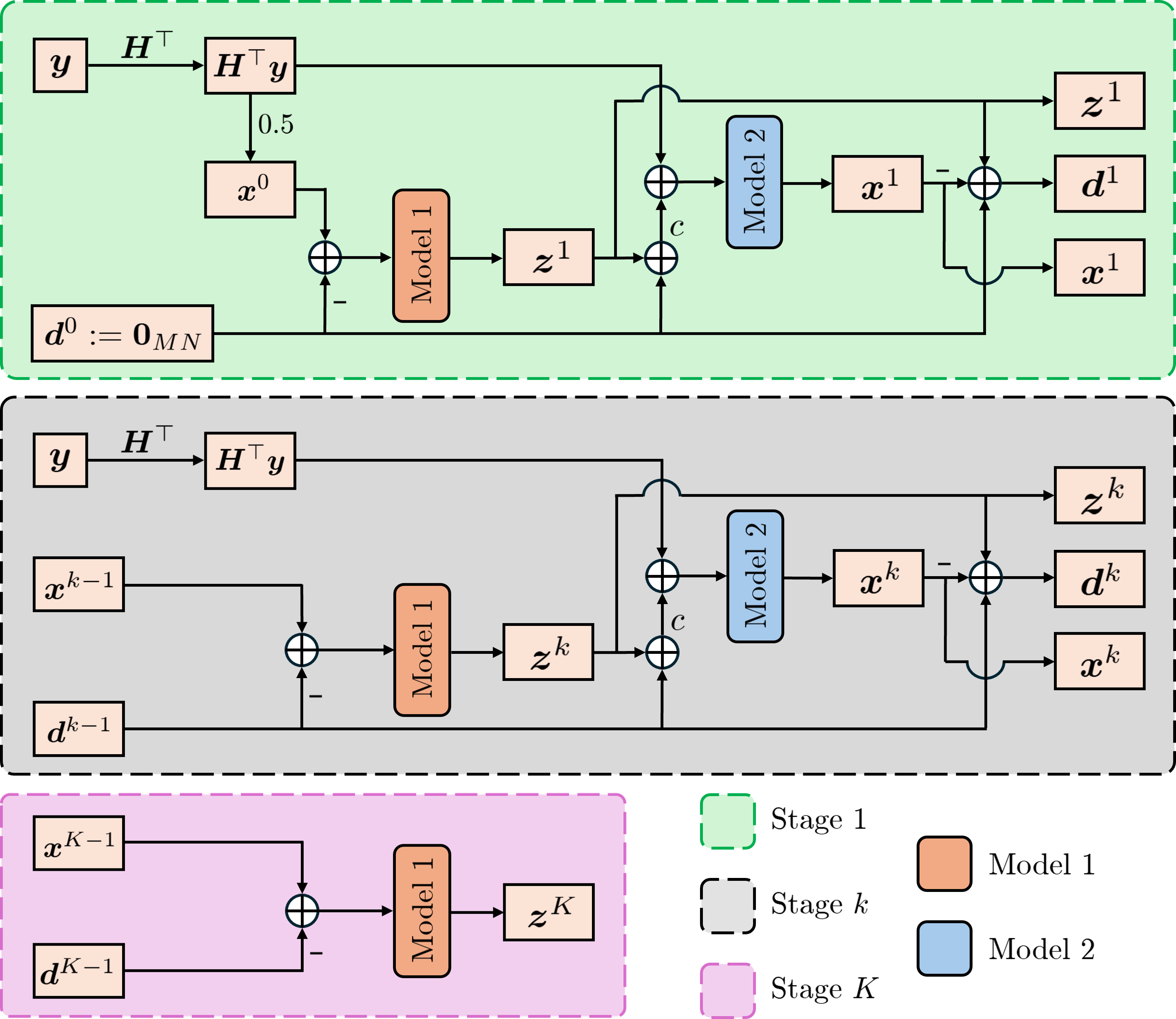}
\caption{Deep unfolding architectures of the proposed COSUP convex algorithm (cf. Section \ref{sec:algodesign}). 
For the intermediate stages $k=2,\dots,K-1$, the architecture is deployed based on the three ADMM closed-form solutions (cf. Section \ref{sec:algodesign}) for updating the primal variables $(\bz,\bx)$ and the dual variable $\bd$.
Accordingly, the intermediate architecture $k$ is augmented to incorporate the initialization for Stage 1, and simplified to compute only the TISR result $\bz$ for Stage $K$.
Model 1 accounts for the Transformer-driven proximal operator for the primal update of $\bz$.
Model 2 derived from the lightweight-driven Woodbury matrix identity accounts for the primal update of $\bx$.
As for the dual update of $\bd$, it is already inherently implemented by the deployed network architectures.}
\label{fig:stage_k}
\end{figure}

To get a sense, we start with the simplest case, which is the dual update \eqref{eq:dual_update}.
The deep architecture deployed in Figure \ref{fig:stage_k}, in which $K$ is the maximum number of ADMM iterations, returns the updated dual variable $\bd$ based on the simple arithmetic operations, i.e., $\bd:= \bd - \bx + \bz$ (cf. \eqref{eq:dual_update}), for the initial stage (i.e., Stage $1$) and the intermediate stages $k=2,\dots,K-1$.
For the final stage (i.e., Stage $K$), the ADMM just needs to return the updated result of $\bz$ (i.e., the TISR result; cf. \eqref{eq:regularized_problem}).
So, there is no need to update the dual variable $\bd$ (or the primal variable $\bx$) in Stage $K$, as illustrated in Figure \ref{fig:stage_k}.
Similar idea will be applied to unfold the ADMM closed-form solutions \eqref{eq:z_update_problem} and \eqref{eq:x_update_closed} using deep architectures for the primal updates of $\bz$ and $\bx$.

\begin{figure}[t]
\centering
\includegraphics[width=1.0\linewidth]{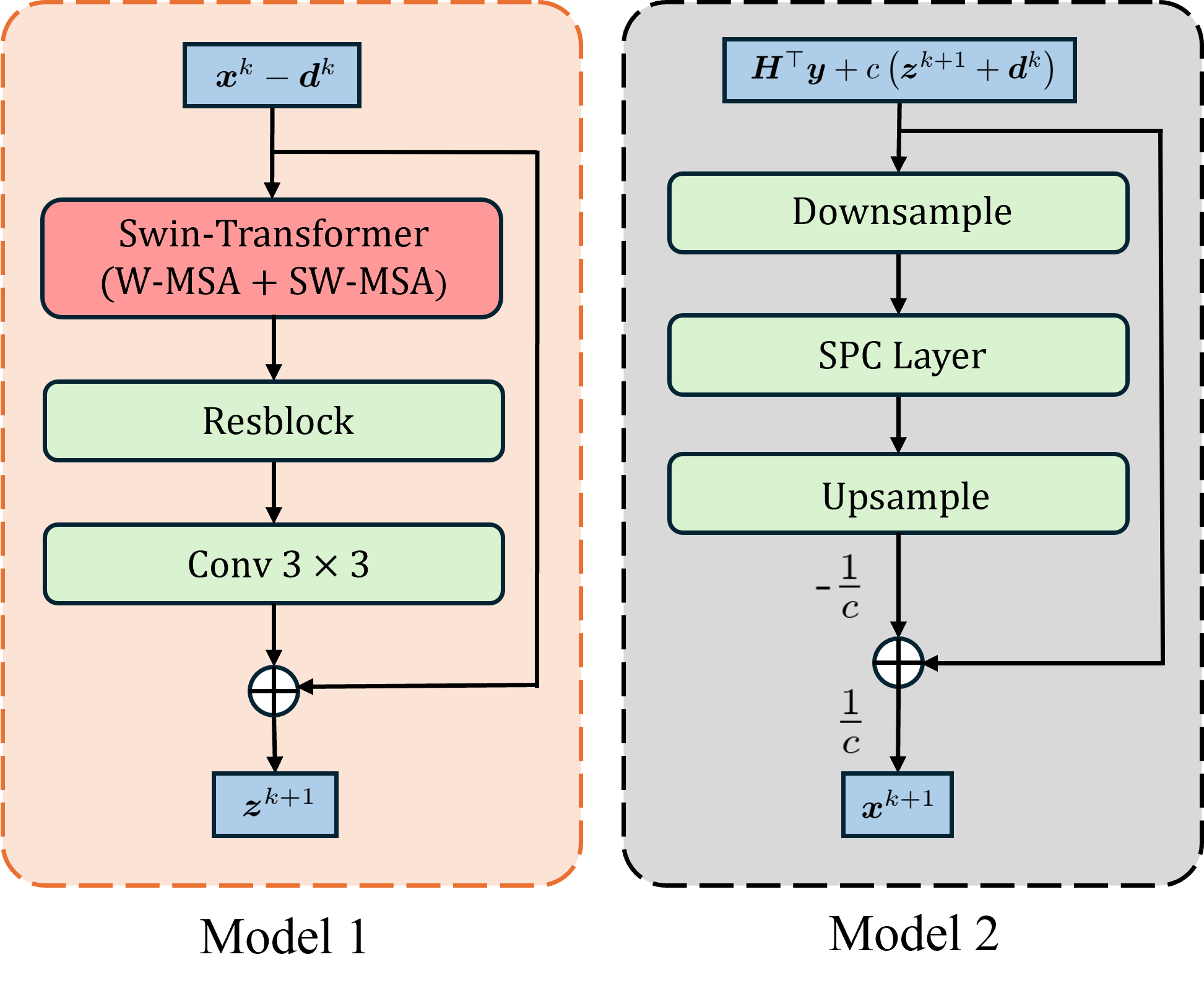}
\caption{Model 1 corresponds to the primal update of $\bz$, and is designed based on the physical meaning of the proximal operator.
The Swin-Transformer, together with the embedded W-MSA and SW-MSA modules, are further illustrated in Figure \ref{fig:swinT}.
Model 2 then implements the primal update of $\bx$, and is designed based on the Woodbury matrix inversion lemma \cite{CVXbookCLL2016}.
The symmetric fully connected (SFC) layer contains a learnable symmetric matrix $\bm\Phi$.
Both models serve as the elementary blocks in the overall deep unfolding network of COSUP algorithm, as depicted in Figure \ref{fig:stage_k}.}
\label{fig:model12}
\end{figure}

Next, we design the deep unfolding network for the update of $\bz$.
By \eqref{eq:z_update_problem}, we have $\bz^{k+1}=\operatorname{prox}_{\frac{\lambda}{c} f}\!\big(\bx^{k}-\bd^{k}\big)$.
According to the definition of proximal mapping \cite{parikh2014proximal}, ``$\textrm{prox}_{g}(\bv)$'' is a denoising operator based on the prior $g$, where $\bv$ is considered as the noisy image, and $\textrm{prox}_{g}(\cdot)$ returns a denoised image with property specified by $g$.
Therefore, the primal update $\bz^{k+1}$ can be obtained by denoising the image ``$\bx^{k}-\bd^{k}$'' using a deep denoising network designed based on the self-similarity prior $\frac{\lambda}{c} f$, as illustrated in Figure \ref{fig:model12}.
This is referred to as the deep plug-and-play strategy \cite{COS2A,zhang2021plug}.
Accordingly, the noisy image $\bx^{k}-\bd^{k}$, formed at each stage using the deep unfolding network in Figure \ref{fig:stage_k}, will need to be fed into a proximal denoising network designed based on the convex self-similarity regularizer $f(\cdot)$ defined in \eqref{eq:ss_phi}.

\begin{figure}[t]
\centering
\includegraphics[width=1.0\linewidth]{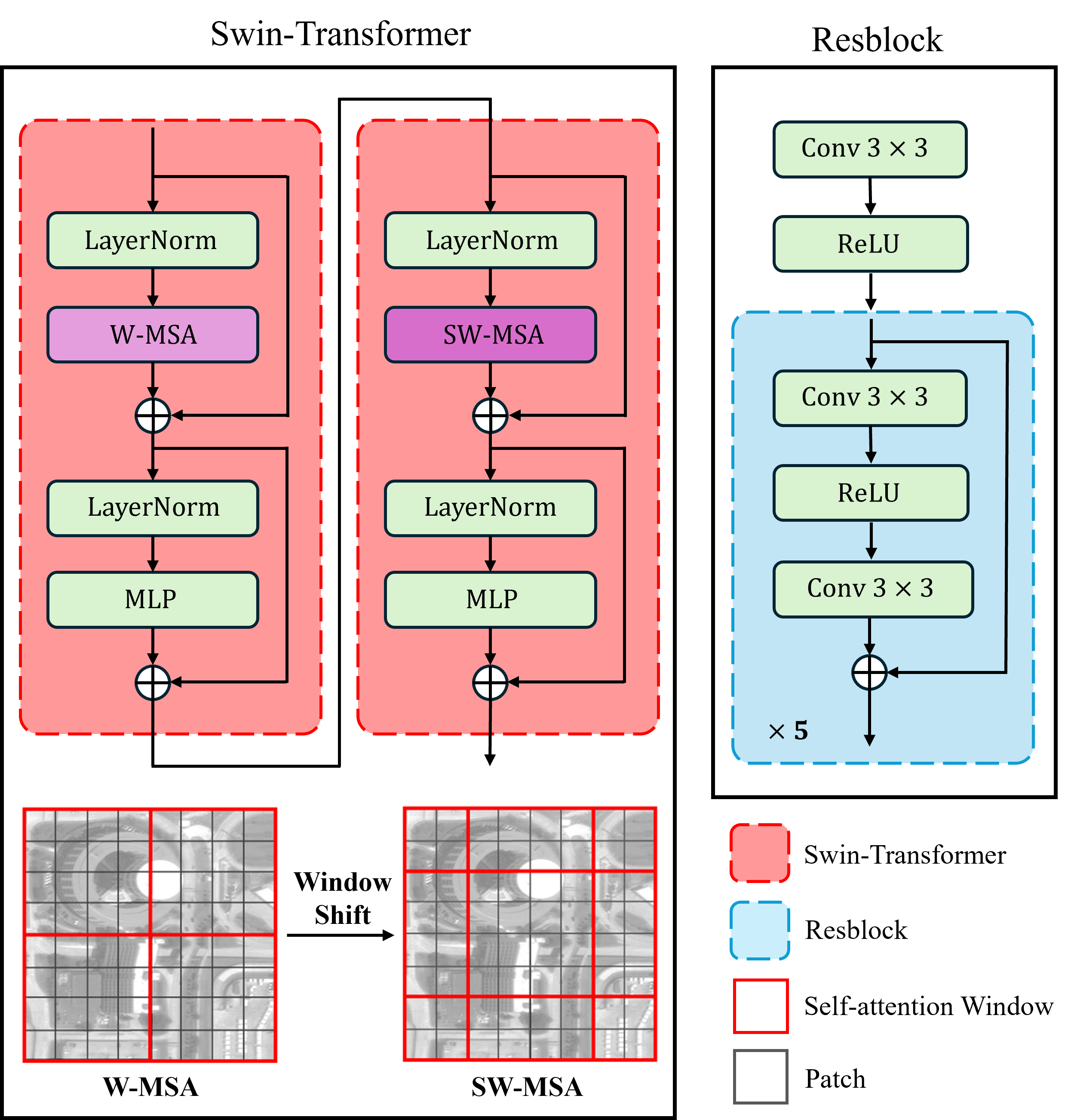}
\caption{Detailed architectures of the Swin-T and the ResBlock, and the graphical illustration of the window-based multi-head self-attention (W-MSA) and shifted window multi-head self-attention (SW-MSA).
The joint adoption of W-MSA and SW-MSA facilitates the learning of both local features and cross-window dependencies.}
\label{fig:swinT}
\end{figure}

The proximal denoising network is designed as model 1 in Figure \ref{fig:model12}, as explained below.
As the residual-in-residual architecture is known to be a simple yet effective denoiser \cite{zhang2017beyond}, it inspired the design of model 1.
The Residual Block (ResBlock) used in model 1 is detailed in Figure \ref{fig:swinT}.
To further promote the image self-similarity, the inner residual network is empowered by the Swin-Transformer \cite{liu2021swin}(Swin-T; cf. Figure \ref{fig:model12}) trained using the self-similarity loss function defined in \eqref{eq:ss_phi}.
To be more specific, the Swin-T first maps the input image into the feature space, followed by a hierarchical structure composed of Swin Blocks. 
Within each Swin Block, the features sequentially pass through Layer Normalization (LayerNorm) and either Window-based Multi-Head Self-Attention (W-MSA) or Shifted Window-based Multi-Head Self-Attention (SW-MSA), as detailed in Figure \ref{fig:swinT}.
W-MSA computes attention within fixed windows to focus on local features, while SW-MSA shifts the window by half its size to enable cross-window feature interaction, as graphically illustrated in Figure \ref{fig:swinT}.
The attention module is followed by residual connections and a Multi-Layer Perceptron (MLP), with LayerNorm inserted to stabilize the model training.
This cross-window feature interaction echos the self-similarity architecture, and makes it an effective and generalizable proximal operator in the deep unfolding framework.
The Swin-T is pretrained with the training details given in Appendix \ref{appdx:trainSwinT}.
The pretrained Swin-T is then plugged into the overall deep unfolding network (cf. Figure \ref{fig:stage_k}) to guide the training of the remaining parts (i.e., those parts other than Swin-T), thereby allowing Swin-T to effectively preserve the self-similarity
structural prior inside the deep network.

Finally, we design the deep architecture to unfold the closed-form solution \eqref{eq:x_update_closed} for the primal update of $\bx$.
In \eqref{eq:x_update_closed}, we need to address the inversion of an $MN\times MN$ matrix. 
Since $\bH^{\top}\bH$ must be positive semidefinite (PSD) for any $\bH$, the eigenvalues of $\bH^{\top}\bH$ must be no less than zero.
This implies that $\bH^{\top}\bH + c\bI$ has eigenvalues no less than $c>0$ (strictly greater than zero), implying that $\bH^{\top}\bH + c\bI$ is positive definite (PD) and hence invertible.
This enables the adoption of the Woodbury matrix identity \cite{CVXbookCLL2016} to reduce the dimensionality of matrix inversion from $MN\times MN$ to $2mn\times 2mn$.
%
Specifically, \eqref{eq:x_update_closed} can be equivalently reformulated as
\begin{equation} 
\label{eq:x_update_final}
\bx^{k+1} 
= 
\frac{1}{c} \left( \bI - \frac{1}{c} \bH^\top \boldsymbol{\Phi} \bH \right)
\left( \bH^\top \by + c (\bz^{k+1} + \bd^k) \right),
\end{equation}
where $\boldsymbol{\Phi} \triangleq \left( \bI + \frac{1}{c} \bH \bH^\top \right)^{-1} \in \mathbb{R}^{2mn \times 2mn}$ corresponds to the $2mn\times 2mn$ matrix inversion.
Thus, the remaining task is to implement \eqref{eq:x_update_final}.

To this end, we need to design model 2 to implement the operator ``$\frac{1}{c} \left( \bI - \frac{1}{c} \bH^\top \boldsymbol{\Phi} \bH \right)$'' (cf. Figure \ref{fig:model12}), and need to form the input image ``$\mathcal{IM}\triangleq \bH^\top \by + c (\bz^{k+1} + \bd^k)$'' (cf. Figure \ref{fig:stage_k}) to be fed into the model 2 for implementing \eqref{eq:x_update_final}.
The deep unfolding network has already been deployed to form the input image $\mathcal{IM}$, as can be observed from Figure \ref{fig:stage_k}.
The model 2 is also designed in Figure \ref{fig:model12}, where the shortcut connection corresponds to the identity matrix in \eqref{eq:x_update_final}.
Since $\bH$ in \eqref{eq:x_update_final} is essentially an image downsampling operator, the physical meaning of $\bH^\top$ is the image upsampling operator, as designed in Figure \ref{fig:model12}.
As for the matrix $\boldsymbol{\Phi}$ in \eqref{eq:x_update_final}, we recall the fact that a matrix can be implemented by a fully connected layer.
However, as $\boldsymbol{\Phi}$ is defined to be a symmetric matrix, it is implemented by a symmetric fully connected (SFC) layer \cite{COS2A} in the deep unfolding, as illustrated in Figure \ref{fig:model12}.
Therefore, we have completed the design of the deep unfolding network for implementing the COSUP algorithm, whose training strategy is described in Appendix \ref{appdx:trainSwinT}.
The superiority of the proposed COSUP algorithm will be experimentally demonstrated.

\section{Experimental Results and Discussion}\label{sec:experiment}

This section presents a comprehensive evaluation of the proposed COSUP algorithm for the TISR task. 
Section~\ref{sec:experimental setting} introduces the overall experimental setup, including the adopted datasets, degradation models, baseline methods, and computer facilities. 
Section~\ref{sec:simulation data} conducts experiments on simulated data to verify the reconstruction quality of COSUP through quantitative metrics and visual comparisons. 
In Section~\ref{sec:real data (SPOT-5)}, we evaluate the method on real SPOT-5 Très Haute Résolution (THR)-mode data, consisting of two LR images (cf. Figure \ref{fig:TISRillustration}), and compare the COSUP-reconstructed HR image against the official CNES SWH Level~2A-Carto (supermode) product, using no-reference image quality metrics and visual analysis.
We also conduct an ablation study to show the critical role of the convex self-similarity prior in yielding high-fidelity supermode imaging results.
%

\subsection{Experimental Setting}\label{sec:experimental setting}

To train and test the proposed COSUP model and related models, we collect HR satellite remote sensing imagery datasets, including GeoEye-1 data (0.58\,m GSD), CBERS-4A data (2.0\,m GSD), and PlanetScope data (3.0\,m GSD) \cite{pandata}, as detailed in Table~\ref{tab:dataset_comparison}. 
These images span regions such as Houston (USA), Mato Grosso (Brazil), and Manta (Ecuador), with acquisition dates ranging from 2016 to 2022. 
The datasets cover a broad spectrum of spatial resolutions and terrain characteristics, providing multi-scale information that enables the simulation of diverse imaging conditions and the training of models with strong generalization ability. 
As will be seen in Section~\ref{sec:real data (SPOT-5)}, even if the training datasets (cf. Table \ref{tab:dataset_comparison}) are completely independent of the SPOT-5 dataset, the superiority of the proposed COSUP algorithm on real SPOT-5 testing data demonstrates its outstanding generalization ability.
To emulate a PAN sensor, each RGB image from the PlanetScope satellite is converted into a single-channel HR PAN image $\bz$ using the NTSC (National Television System Committee) luminance weights $(0.30R+0.59G+0.11B)$\cite{poynton2012digital}, where $\bz$ will be used to simulate the two LR images (cf. Figure \ref{fig:TISRillustration}) and will serve as the ground truth of the TISR problem.

\begin{table}[t]
\centering
\caption{The HR PAN datasets are available \cite{pandata} for generating the LR pairs (i.e., $\by_1$ and $\by_2$) in our TISR experiments.
The data are acquired from GeoEye-1, CBERS--4A, and PlanetScope satellites, whose spectral ranges (wavelength) are specified in $\mu$m.}
\label{tab:dataset_comparison}
\renewcommand{\arraystretch}{1.1}
\setlength{\tabcolsep}{8pt}
\begin{tabular}{|c|c|c|c|}
\hline
Satellite & GeoEye-1 & CBERS--4A & PlanetScope \\
\hline
GSD (m) & 0.58 & 2.0 & 3.0 \\
\hline
Wavelength & 0.45--0.80 & 0.51--0.85 &
\begin{tabular}{@{}c@{}}
B: 0.46--0.51 \\
G: 0.54--0.59 \\
R: 0.65--0.68
\end{tabular} \\
\hline
Site & Houston & Mato Grosso & Manta \\
Country & USA & Brazil & Ecuador \\
Year & 2017 & 2021, 2022 & 2016 \\
Image Size & $19584\times19584$ & $27656\times27656$ & $7379\times6847$ \\
\hline
\end{tabular}
\end{table}

To generate simulation data under the TISR setting, the collected large-scale HR PAN images are divided into nonoverlapping $512 \times 512$ patches, which serve as the ground truth $\bz$. 
As many experiments are designed based on the Wald protocol \cite{COCNMF,PRIME}, it is adopted in this paper.
Specifically, the first $256 \times 256$ LR image $\by_{1}$ is produced by convolving each HR patch with a $7 \times 7$ Gaussian blurring kernel with the variance $v = 0.65$, followed by $2 \times 2$ downsampling, to better approximate the real sensor point spread function (PSF) \cite{rs6087491,217219}.
The twin $256 \times 256$ LR image $\by_{2}$, corresponding to a half-pixel displacement relative to $\by_{1}$, is obtained by shifting the HR patch one pixel leftward and one pixel downward with nearest-pixel boundary extension, after which the same procedure for simulating $\by_{1}$ is applied to simulate $\by_{2}$.
As a result, the LR pair exhibits a consistent half-pixel offset in both horizontal and vertical directions, and will be fed into TISR models to estimate the HR image $\widehat{\bz}$.
We will discuss how to measure the similarity between $\widehat{\bz}$ and its ground truth ${\bz}$ in Section \ref{sec:simulation data}, thereby quantitatively comparing related TISR models.

The collected HR images described above covering four representative land types \cite{HyperKING}, including city, farm, coastline, and mountain.
Each scene is cropped into nonoverlapping patches, and input pairs \((\by_{1}, \by_{2})\) are synthesized following the procedure described above.
The training and testing patches are drawn from geographically disjoint regions. 
Initially, this yields 2200 samples for training, 300 for validation, and 105 for testing, all based on the ideal shift configuration (i.e., 0.5-pixel shift configuration).
To further enhance the model generalization ability against real-world misregistration noise (see Figure \ref{fig:Shift} for the nonstandard real subpixel offsets), we enriched the training dataset by adding 1200 additional samples with random subpixel offsets (widely ranging from 0 to 1 pixel with a step size of 0.1), together with 50 additional validation samples also with random subpixel offsets, as will be further detailed in Section~\ref{sec:ablation}. 
Consequently, the final dataset consists of 3400 training samples and 350 validation samples, which better meet the real-world subpixel offsets (i.e., nonideal shift configuration).
%
During the training phase, we apply typical data augmentation tricks, including random horizontal/vertical flips and rotations.
All augmentation operations are applied consistently to each HR patch and its corresponding LR pair so that the subpixel displacement between $\by_{1}$ and $\by_{2}$ is preserved.

Some pretraining details are provided in Appendix~\ref{appdx:trainSwinT}.
In the pretraining phase of Swin-T, the network is optimized using Adam with $\beta_1 = 0.9$, $\beta_2 = 0.999$, and $\epsilon = 10^{-8}$.
The initial learning rate is set to $5 \times 10^{-4}$ and is halved whenever the validation PSNR does not improve for five consecutive epochs, with a minimum learning rate of $10^{-7}$.
Training is conducted for 110 epochs with a batch size set as 6.
After the pretraining phase, the COSUP unfolding framework is configured with $K:=3$ stages.
The penalty parameter $c$ is learnable, with its initialization given by $2$.
This design allows our COSUP algorithm to automatically learn/optimize the parameter via ADAM optimizer during training \cite{kingma2014adam}.
%
%
All variables are initialized as zeros. 
%
Other configurations, including optimizer, learning rate, and scheduling strategy, follow the same pretraining settings as described above.

All experiments are conducted under identical setups for all models. The implementation is based on MATLAB R2023b, Python 3.10.12, and PyTorch 2.7.1 (CUDA 12.6), and executed on a workstation equipped with an Intel Core i9-10900K CPU (3.70 GHz), 125 GB RAM, and an NVIDIA GeForce RTX 3090 GPU.

\subsection{Quantitative and Qualitative Performance Evaluation}\label{sec:simulation data}

To have a fair comparison among COSUP and existing baselines, we quantitatively evaluate the similarity between the super-resolved image (output of each method) and the ground-truth HR image available in the aforementioned simulated dataset, using two widely adopted full-reference image quality assessment metrics \cite{CODEIF,CODE}, including peak signal-to-noise ratio (PSNR) and structural similarity index (SSIM).
PSNR measures the pixel-wise error between the reconstructed image and the ground truth, where a higher value indicates that the reconstruction is numerically closer to the ground truth.
SSIM assesses reconstruction quality from the perspectives of luminance, contrast, and structural consistency.
Its values range from 0 to 1, where a higher value suggests that the reconstructed image is more structurally consistent with the ground truth.

\begin{table}[t]
\centering
\caption{Quantitative comparisons among SR methods, including bicubic, IBP \cite{irani1991improving}, BTV \cite{farsiu2004fast}, MAN \cite{wang2024man}, SUM \cite{molini2019deepsum}, and the proposed COSUP algorithm, across diverse landscapes.
Boldfaced numbers indicate scene-specific best performances.}
\label{tab:psnr_ssim_by_scene}
\renewcommand{\arraystretch}{0.95}
\setlength{\tabcolsep}{6pt}
\resizebox{1.0\linewidth}{!}{
\begin{tabular}{c l c c c}
\hline
Scene & Method & PSNR ($\uparrow$) & SSIM ($\uparrow$) & Time (sec.)\\
\hline
\multirow{6}{*}{\makecell[c]{Farm\\(30)}} 
  & Bicubic  & 36.35 & 0.9200 & \textbf{0.051}\\
  & IBP      & 39.72 & 0.9535 & 6.420\\
  & BTV      & 39.52 & 0.9507 & 6.442\\
  & MAN      & 43.91 & 0.9786 & 0.381\\
  & SUM      & 40.79 & 0.9740 & 0.313\\
  & COSUP    & \textbf{44.92} & \textbf{0.9869}&\textbf{0.051} \\
\hline
\multirow{6}{*}{\makecell[c]{City\\(30)}} 
  & Bicubic  & 30.22 & 0.9145 & 0.063\\
  & IBP      & 32.20 & 0.9369 & 6.588 \\
  & BTV      & 32.13 & 0.9354 & 6.306 \\
  & MAN      & 37.52 & 0.9751 &0.385\\
  & SUM      & 34.38 & 0.9640 &0.320 \\
  & COSUP    & \textbf{39.56} & \textbf{0.9858}&\textbf{0.059} \\
\hline
\multirow{6}{*}{\makecell[c]{Mountain\\(25)}} 
  & Bicubic  & 39.31 & 0.9533 & \textbf{0.046}\\
  & IBP      & 43.50 & 0.9751 &6.458\\
  & BTV      & 43.20 & 0.9730 &6.370 \\
  & MAN      & \textbf{48.03} &0.9890 &0.392 \\
  & SUM      & 43.27 & 0.9838 &0.361\\
  & COSUP    &47.47  & \textbf{0.9914}& 0.052 \\
\hline
\multirow{6}{*}{\makecell[c]{Coastline\\(20)}} 
  & Bicubic  & 37.17 & 0.9439 &0.062\\
  & IBP      & 39.28 & 0.9532 & 6.491\\
  & BTV      & 39.12 & 0.9510 &7.396\\
  & MAN      & 43.64 & 0.9816 & 0.393\\
  & SUM      & 40.62 & 0.9744 &0.363\\
  & COSUP    & \textbf{44.93} & \textbf{0.9860}& \textbf{0.054} \\
\hline\hline
\multirow{6}{*}{\makecell[c]{Average\\}} 
  & Bicubic  & 35.46 & 0.9309 &0.055\\
  & IBP      & 38.38 & 0.9538 & 6.498\\
  & BTV      & 38.21 & 0.9517 &6.568\\
  & MAN      & 43.01 & 0.9806 & 0.387\\
  & SUM      & 39.52 & 0.9736 &0.336\\
  & COSUP    & \textbf{44.00} & \textbf{0.9857}& \textbf{0.054} \\
\hline
\end{tabular}}
\end{table}

\begin{figure*}[!t]
\centering
\hspace*{-6mm}\includegraphics[width=0.8\textwidth]{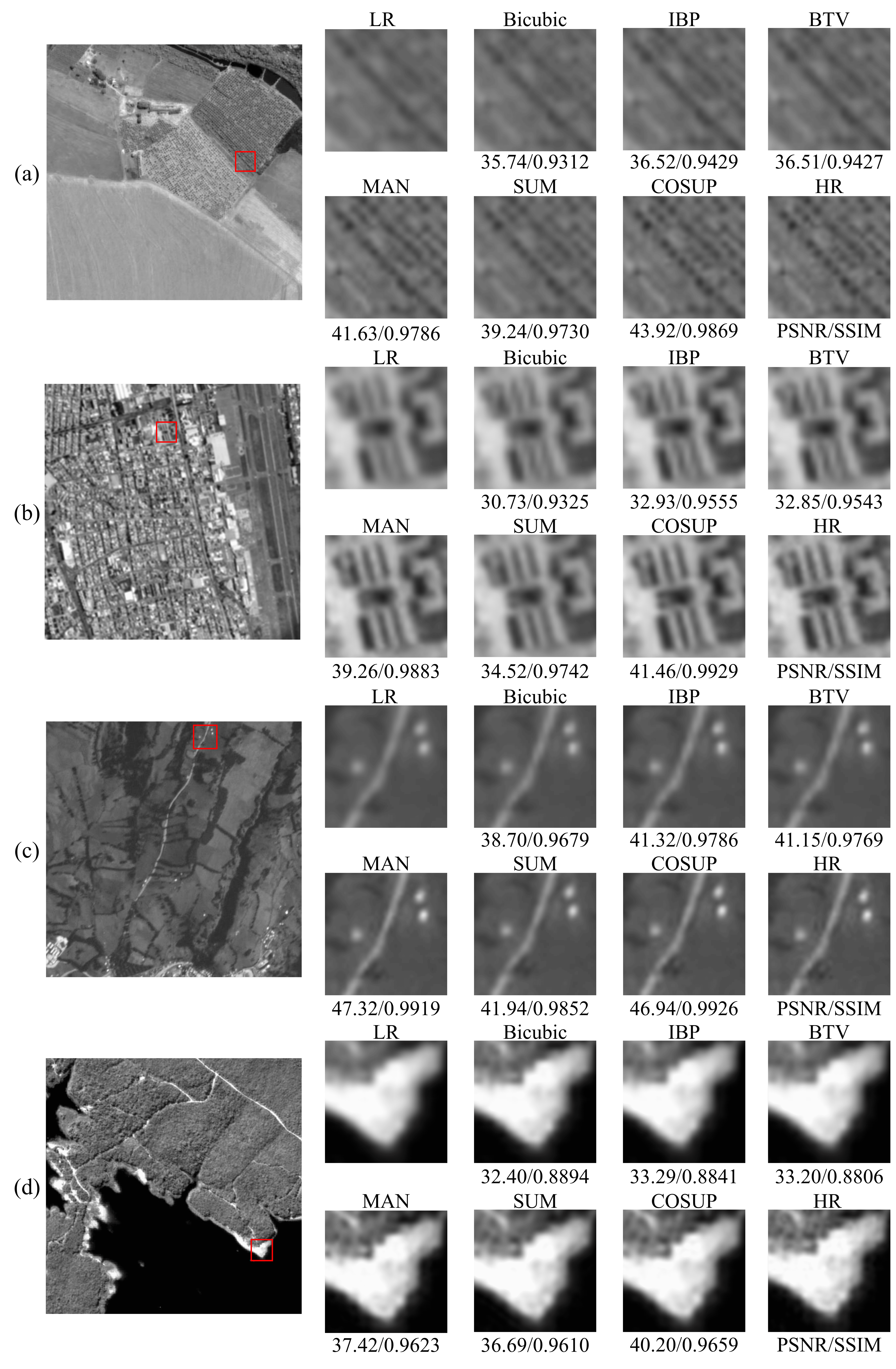}\\[6pt]
\caption{ Visualized SR results across four representative scenes, including (a) farm, (b) city, (c) mountain, and (d) coastline.
The HR image denotes the ground-truth $\bz$, and the LR image is the input $\by_1$, as depicted in Figure \ref{fig:TISRillustration}.
The numbers below each subimage denote the PSNR/SSIM values.
}
\label{fig:simulation_figs}
\end{figure*}

The proposed COSUP algorithm is compared with representative baseline models, and the experimental results across diverse land scapes are summarized in Table \ref{tab:psnr_ssim_by_scene} for quantitative evaluation and Figure \ref{fig:simulation_figs} for qualitative evaluation.
The baseline models include the bicubic method, iterative back projection (IBP) \cite{irani1991improving}, bilateral total variation (BTV) \cite{farsiu2004fast}, multi-scale attention network (MAN)  \cite{wang2024man}, and super-resolution of unregistered multiple images (SUM) \cite{molini2019deepsum}.
IBP, BTV and SUM can address subpixel-shift TISR, while bicubic and MAN were proposed for SR of single image.
We purposely compare with MAN as it is a widely known SISR method published very recently, having drawn considerable attention in related areas.
For bicubic and MAN, they are applied to the LR input image $\by_1$ to obtain the HR results for the subsequent quantitative and qualitative evaluations.

IBP reconstructs the HR image by iteratively back-projecting residuals under the forward degradation model, while BTV adds a bilateral TV regularizer to preserve edges while suppressing noise.
MAN combines multi-scale large-kernel attention and a gated spatial attention unit to capture long-range and local contexts, improving texture reconstruction while keeping computational cost moderate.
SUM learns registration and uses 3D CNN feature fusion to align and aggregate unregistered images, improving detail recovery.
To ensure a fair evaluation, those learning-based methods (i.e., MAN and SUM) are retrained using the same dataset, strictly following the original training settings and hyperparameters described in their respective papers. 
During the training phase, MAN only needs the single input image $\by_1$ without requiring the twin image $\by_2$ (as it is for SISR).
In the testing phase, the SISR methods (i.e., bicubic and MAN) use only one input image $\by_1$ as well. 
All other methods, including IBP, BTV, SUM, and our proposed COSUP, take two subpixel-shift input images $\by_1$ and $\by_2$, under the TISR setting.
Accordingly, the quantitative results are reported in Table~\ref{tab:psnr_ssim_by_scene}.

COSUP achieves the highest PSNR for land types of farm, city, and coastline, while MAN performs best in terms of PSNR for the mountain area.
For SSIM, the proposed COSUP algorithm consistently yields the strongest performance over all the landscapes, showing its outstanding SR capability in preserving the luminance, contrast, and structural properties.
This demonstrates the superiority and effectiveness of the convex self-similarity prior used in the design of COSUP.
MAN performs well because of its multi-scale large-kernel attention scheme and gated spatial units, capturing both long-range structures and local details.
Except for bicubic, the other baselines also show reasonably good results, though occasionally exhibiting unstable performance, as observed in the city and coastline areas.

Figure~\ref{fig:simulation_figs} shows visual reconstruction results for all the methods under test.
Across all the landscapes, COSUP consistently preserves fine details and sharp edges compared to other methods. 
Moreover, COSUP is able to accurately reconstruct narrow features (e.g., buildings, and field boundaries) while maintaining local contrast, echoing the strong SSIM performance in Table~\ref{tab:psnr_ssim_by_scene}.
For example, in the farm scene shown in Figure~\ref{fig:simulation_figs}(a), COSUP reconstructs fine details within the field region, separates dark elements that appear merged in the LR inputs, and preserves the spacing and local contrast among crop rows. 
By contrast, MAN and SUM tend to produce blurred textures and a loss of fine details. 
In the city scene shown in Figure~\ref{fig:simulation_figs}(b), a local zoom on the building region shows that COSUP recovers building contours and small structural details with sharper edges and a geometry closer to the HR reference, whereas competing methods yield softer, less accurate boundaries. 
In Figure~\ref{fig:simulation_figs}(c) and Figure~\ref{fig:simulation_figs}(d), MAN and SUM produce overly smooth outputs, where edges are blurred, high-contrast features are weakened, and bright spots appear broadened, which together obscure fine structures such as mountain roads and sharp shoreline corners. 
COSUP recovers much sharper contours and preserves local texture, yielding well-separated spot-like responses in farmland patches, crisper linear edges along coastlines, and clearer road/detail preservation in mountainous areas, demonstrating the effectiveness of the proposed convex self-similarity regularization strategy (cf. Figure~\ref{fig:simulation_figs}).
Remarkably, the computational time of COSUP is almost as fast as the near-real-time bicubic baseline, thanks to the lightweight architecture design behind the proposed COSUP algorithm.

\subsection{ Nonideal Subpixel Shifts and Model-Order Selection
}\label{sec:ablation}

In Sections \ref{sec:experimental setting} and \ref{sec:simulation data}, two LR images are ideally offset by half a pixel both horizontally and vertically in the testing stage.
Considering that such an exact half-pixel shift would be violated under real-world scenarios (as evidenced in Figure \ref{fig:Shift}), we conduct an experiment to evaluate the performance on the same 105 testing ROIs (cf. Section \ref{sec:experimental setting}) but with nonideal subpixel shifts ranging from 0 to 1 pixel (with a step size of 0.1).
%
%
The corresponding experiment is designed below.
To generate simulation data with nonideal subpixel displacements (i.e., $\by_2$) from $\bz$, the simulation protocol of $2\times 2$ downsampling (cf. Section \ref{sec:experimental setting}) is no longer applicable.
To address this, we customize a new experimental protocol that leverages the original satellite imagery.
In detail, we use GeoEye-1, CBERS-4A, and PlanetScope imagery \cite{pandata} for generating HR images $\bz$ with $5\times 5$ downsampling, and the corresponding LR images $\by_1$ with $10\times 10$ downsampling.
%
%
Consequently, a 1-pixel shift in the original imagery corresponds to a 0.1-pixel shift in the LR domain; similarly, an $n$-pixel shift in the original imagery corresponds to a (0.1$n$)-pixel shift in the LR domain for any integer $n$. 
The ideal case corresponds to $n:=5$ (i.e., half-pixel shift), and the cases with $n\neq 5$ are for nonideal scenarios.
This allows us to control $\by_2$ with any prespecified subpixel offsets.
Using the above customized experimental protocol, we further demonstrate that our dataset enrichment strategy (i.e., adding 1200 additional training samples with nonideal subpixel offsets; cf. Section \ref{sec:experimental setting}) does remarkably upgrade the generalizability of COSUP and its robustness against the nonideal offsets; this will be trickily demonstrated by comparing the two COSUP versions trained with or without the dataset enrichment strategy (the one without enrichment is referred to as the baseline model.

\begin{figure}[t]
    \centering
    \includegraphics[width=0.95\linewidth]{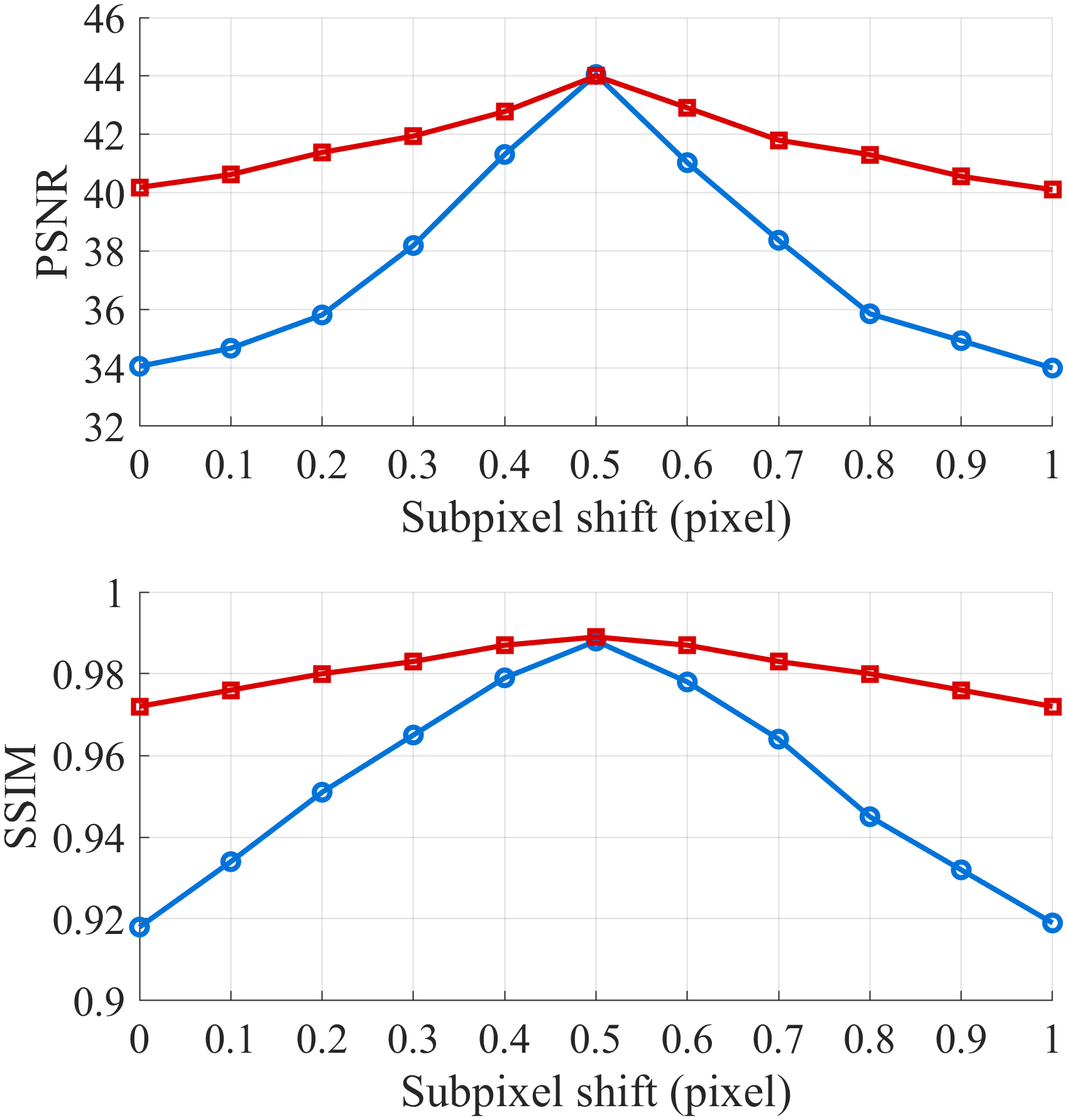}
    \caption{COSUP performances across different subpixel-shift levels, ranging from 0 to 1 pixel with a step size of 0.1.
    The red and blue curves correspond to the two COSUP versions trained with and without the dataset enrichment strategy (cf. Section \ref{sec:experimental setting}), respectively.
    Compared to the baseline model (blue curve), the dataset enrichment strategy (red curve) does significantly upgrade the generalizability of COSUP over diverse nonideal offsets.
    Remarkably, even for the extreme offsets, COSUP still exhibits strong performances with PSNR of about 34dB and SSIM of about 0.92, when the dataset enrichment strategy is not applied (blue curve).
    }
    \label{fig:shift_curve}
\end{figure}

The experimental results are presented in Figure \ref{fig:shift_curve}.
As expected, both the PSNR and SSIM values are always the highest at the 0.5-pixel shift (the ideal case), confirming that the half-pixel offset maximizes the complementary information.
One can see that as the subpixel shift approaches 0 or 1 pixels, the performance of the baseline significantly degrades in both PSNR and SSIM, thereby forming an obvious bell-shaped curve.
In contrast, by forcing COSUP to learn the nonideal offsets during the training phase (i.e., dataset enrichment strategy), the above situations are greatly mitigated (cf. the red curves in Figure \ref{fig:shift_curve}).
Remarkably, the performances of COSUP under extreme subpixel shifts are still better than the performances of most other peer methods under the ideal case of half-pixel shift (cf. averaged performance in Table \ref{tab:psnr_ssim_by_scene}).
These results collectively demonstrate the robustness of COSUP against real-world misregistration, confirming the necessity and effectiveness of the proposed dataset enrichment strategy.

We also experimentally determine the optimal model-order $K$ (i.e., the number of unfolding stages in COSUP).
As evidenced in the existing unfolding techniques for the SR task \cite{9656645,9904907,COS2A}, it is widely recognized that a moderate searching range of $K \leq5$ is sufficient for achieving strong SR performances.
Accordingly, we conduct an analysis for $K \in \{2,3,4,5\}$.
As it turns out, $K=3$ is identified as the optimal model order based on the PSNR/SSIM performances summarized in Table \ref{tab:ablation_K_c}.

\begin{table}[t]
\renewcommand{\arraystretch}{1.15}
\caption{Performance of COSUP for different model orders $K$ (i.e., the number of unfolding stages).}
\label{tab:ablation_K_c}
\centering
\setlength{\tabcolsep}{6pt}
\resizebox{0.85\linewidth}{!}{
\begin{tabular}{c|c c c}
\hline
 & PSNR ($\uparrow$) & SSIM ($\uparrow$) & Time (sec.)\\
\hline
$K=2$   & 43.58 & 0.986  & 0.030 \\
$K=3$   & \textbf{44.00} & \textbf{0.988} & 0.046  \\
$K=4$   & 43.85  & 0.987  & 0.059 \\
$K=5$   & 43.00  & 0.985  & 0.074 \\
\hline

\end{tabular}}
\end{table}

\subsection{Real Data Evaluation}\label{sec:real data (SPOT-5)}

To design the real-data experiments, we remark that the SWH platform of the French Space Agency (CNES) \cite{swh2acarto} only accepts SPOT-5 THR image pairs downloaded from the SWH REGARDS catalog \cite{spot5data}, not supporting other data sources acquired by other satellites. 
Consequently, other satellites that require supermode technology (e.g., FORMOSAT-8 \cite{formosat-8}) cannot benefit from this service.
Moreover, the mechanism of the supermode algorithm behind the platform has not been openly available, making it difficult to adapt the method for different satellites.
Therefore, the only way to compare with the official supermode product is to feed the SPOT-5 THR image pairs into the platform and COSUP, thereby comparing the quality of the HR images returned by these methods using no-reference image quality metrics; note that there is no official ground truth of the HR image for computing PSNR.

\begin{figure}[t] 
\centering 
\includegraphics[width=1.0\linewidth]{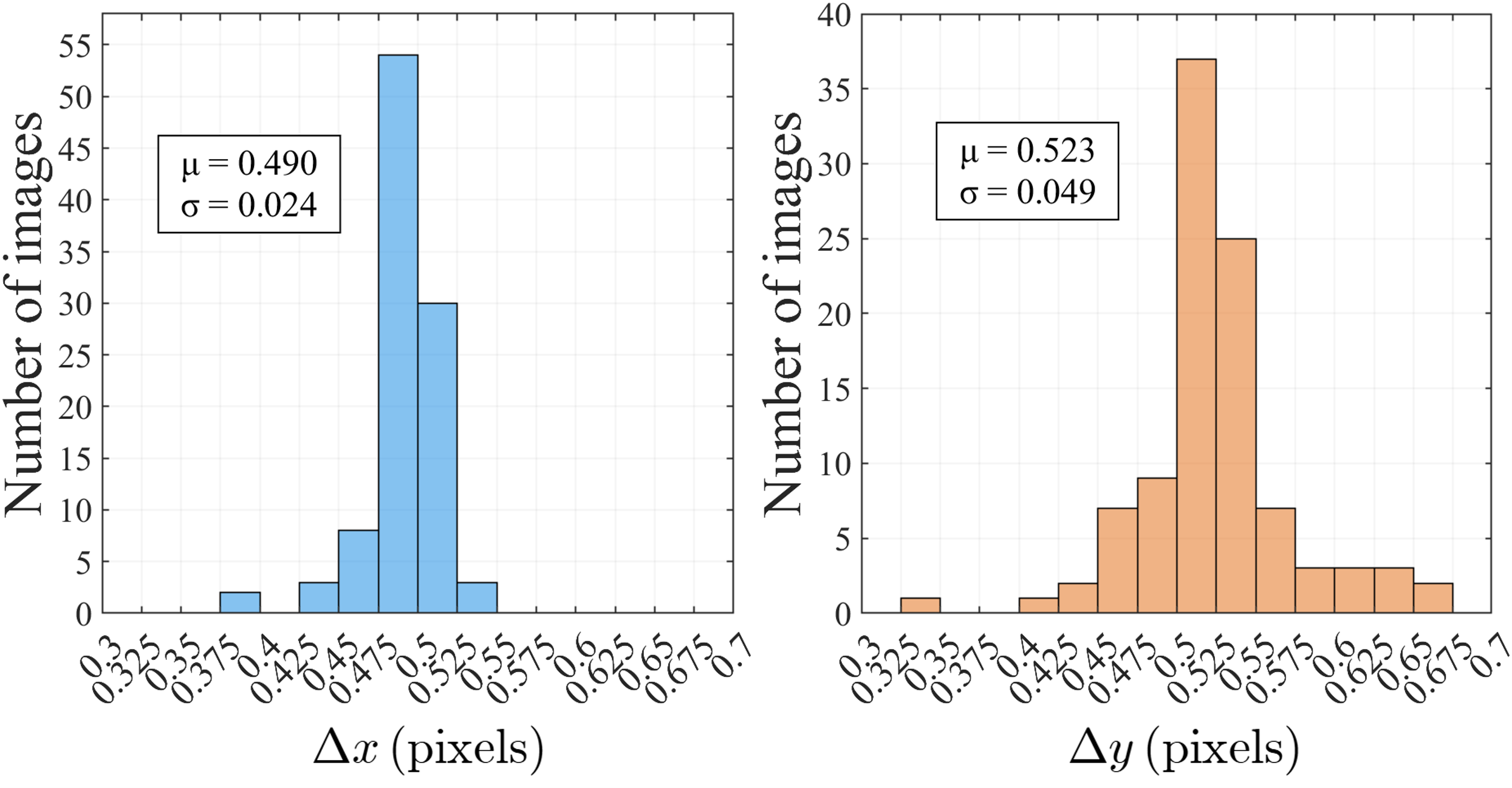} 
\caption{In real-world situation of TISR, the two LR images $\by_1$ and $\by_2$ would not offset exactly by 0.5 pixels both horizontally and vertically (i.e., $\triangle x=\triangle y=0.5$).
This figure displays the distribution and statistics of the subpixel-shift levels \cite{guizar2008efficient} along the horizontal direction ($\triangle x$) and along the vertical direction ($\triangle y$) for the 100 SPOT-5 LR image pairs (cf. Table \ref{tab:spot5_coverage}).
Insets report the mean $\mu$ and standard deviation $\sigma$ for each direction.}
\label{fig:Shift} 
\end{figure}

To be more specific, the real data used in this study were obtained from the SWH REGARDS catalog established by CNES \cite{spot5data}, which archives and disseminates the complete SPOT-5 image collections. 
From this platform, we retrieved paired PAN images acquired in the THR mode of SPOT-5, denoted as SWH-1A products. 
Each image has a ground sampling distance of 5\,m, and the two images within a pair exhibit a roughly half-pixel displacement.
There are in total 100 representative samples selected, as summarized in Table~\ref{tab:spot5_coverage}.
We remark that the real-data situation is more challenging as the subpixel shifts would not be spatially uniform, as illustrated in Figure~\ref{fig:Shift}, where the true subpixel-shift levels are estimated using the embedded MATLAB function ``dftregistration'' \cite{guizar2008efficient}.
The collected LR SPOT-5 paired images can subsequently be submitted to the official SWH-2A-Carto service for TISR processing.
The service first generates the orthorectified and georeferenced SWH Level~2A-Carto product.
When a SPOT-5 THR pair is provided, the system additionally performs supermode fusion and a deconvolution operation to complete the TISR processing, ultimately producing the desired HR supermode image with 2.5\,m resolution \cite{swh2acarto}.

Following the setup in Section~\ref{sec:experimental setting}, we downloaded multiple pairs of SPOT-5 THR images from the SWH platform together with their corresponding supermode HR products.
The original paired LR images were cropped into typical $256 \times 256$ images, from which 100 representative samples were selected with diverse land-cover types (i.e., farm, city, mountain, and coastline) in different countries, as summarized in Table~\ref{tab:spot5_coverage}.
To ensure consistency in scene alignment, the HR supermode outputs were first geometrically aligned based on geographic coordinates, and subsequently cropped into the corresponding $512 \times 512$ HR images to serve as the baseline, for subsequent quantitative and visual comparison.
As for the proposed COSUP, following the comparison protocol in Section~\ref{sec:simulation data}, it accepts the above paired LR images $(\by_1,\by_2)$ to obtain the target HR images, which (and the aforementioned baseline) are evaluated using no-reference quality metrics.

Specifically, we evaluate the results of COSUP and the baseline supermode product using the natural image quality evaluator (NIQE) \cite{mittal2013niqe} and the perceptual index (PI) \cite{chang2015pi}.
NIQE is derived from Natural Scene Statistics (NSS) and evaluates image quality by quantifying the deviation of the statistical distribution of the tested image from those of high-quality natural images.
A lower NIQE score indicates that the tested image more closely adheres to natural statistical characteristics, reflecting reduced distortion and improved visual quality. 
PI integrates multiple no-reference quality indicators to approximate the human visual system’s subjective perception of image quality, with lower values likewise indicating higher perceptual quality \cite{blau2018pirm}.
To highlight performance variations across different land covers, we report the mean values of NIQE and PI for each of the four scene categories in Table \ref{tab:no_ref_metrics}.
One can clearly see that the proposed COSUP algorithm significantly outperforms the official supermode product for all the landscapes (cf. Table \ref{tab:no_ref_metrics}).

\begin{table}[t]
\centering
\caption{SPOT-5 Level~1A panchromatic (PAN) scenes used in our real-data evaluation are summarized in this table.
All the SWH Level~1A pair images are downloaded from \cite{spot5data}.}
\label{tab:spot5_coverage}
\renewcommand{\arraystretch}{1.12}
\setlength{\tabcolsep}{6pt}
\makebox[\linewidth][c]{%
\resizebox{0.98\linewidth}{!}{%
\begin{tabular}{|c|c|c|c|c|}
\hline
\multicolumn{1}{|c|}{Scene Type} & \multicolumn{1}{c|}{Site} & \multicolumn{1}{c|}{Country} & \multicolumn{1}{c|}{Year} & \multicolumn{1}{c|}{Image Size} \\
\hline
\makecell[c]{Farm\\(30)}    & Willows    & USA       & 2009 & $12000\times12000$ \\ \hline
\makecell[c]{City\\(25)}    & Daejeon    & South Korea     & 2009 & $12000\times12000$ \\ \hline
\multirow{2}{*}{\makecell[c]{Mountain\\(25)}}  & Tuba City  & USA       & 2007 & $12000\times12000$ \\ \cline{2-5}
                           & Rason      & DPRK      & 2004 & $12000\times12000$ \\ \hline
\multirow{2}{*}{\makecell[c]{Coastline\\(20)}} & Jurien Bay & Australia & 2006 & $12000\times12000$ \\ \cline{2-5}
                           & Gualala    & USA       & 2007 & $12000\times12000$ \\ \hline
\end{tabular}%
}
}
\end{table}

\begin{table}[!t]
\renewcommand{\arraystretch}{1.2}
\caption{Quantitative evaluations on real SPOT-5 satellite data across diverse land types.
The quantitative results of the CNES supermode imaging product, COSUP-nr (nr: no regularization), COSUP-TV, and COSUP, are demonstrated in terms of no-reference image quality metrics, i.e., NIQE and PI.
Boldfaced numbers indicate scene-specific best performances.}
\label{tab:no_ref_metrics}
\centering
\setlength{\tabcolsep}{6pt}
\resizebox{1.0\linewidth}{!}{
\begin{tabular}{c|>{\raggedright\arraybackslash}p{1.4cm}|c c c}
\hline
Scene & \multicolumn{1}{c|}{Method} & NIQE ($\downarrow$) & PI ($\downarrow$) & Time (sec.) \\
\hline
\multirow{4}{*}{\makecell[c]{Farm\\(30)}} 
 & Supermode & 4.713 & 3.520 & -- \\
 & COSUP-nr   & 4.931 & 3.155 & 0.039 \\
 & COSUP-TV   & 5.853 & 4.598 & 0.041 \\
 & COSUP      & \textbf{3.624} & \textbf{2.610} & 0.050 \\
\hline
\multirow{4}{*}{\makecell[c]{City\\(25)}} 
 & Supermode & 5.600 & 3.659 & -- \\
 & COSUP-nr  & 5.882 & 3.519 & 0.042 \\
 & COSUP-TV   & 6.890 & 4.947 & 0.042 \\
 & COSUP      & \textbf{4.631} & \textbf{2.903} & 0.052 \\
\hline
\multirow{4}{*}{\makecell[c]{Mountain\\(25)}} 
 & Supermode &  4.657 & 3.616 & -- \\
 & COSUP-nr  & 5.893 & 3.947 & 0.040 \\
 & COSUP-TV   & 6.038 & 4.923 & 0.042 \\
 & COSUP      & \textbf{4.207} & \textbf{3.171} & 0.052 \\
\hline
\multirow{4}{*}{\makecell[c]{Coastline\\(20)}}  
 & Supermode &  4.365 & 3.058 & -- \\
 & COSUP-nr   & 4.919 & 3.039 & 0.042 \\
& COSUP-TV   & 5.579 & 4.214 & 0.044 \\
 & COSUP      & \textbf{3.515} & \textbf{2.310} & 0.054 \\
\hline\hline
\multirow{4}{*}{\makecell[c]{Average\\}}  
 & Supermode &  4.851 & 3.486 & -- \\
 & COSUP-nr   & 5.407 & 3.421 & 0.041 \\
 & COSUP-TV   & 6.090 & 4.671 & 0.042 \\
 & COSUP      & \textbf{3.999} & \textbf{2.764} & 0.052 \\
\hline
\end{tabular}}
\end{table}

\begin{figure}[!t]
\centering
\hspace*{-1.7mm}\includegraphics[width=1.01\linewidth]{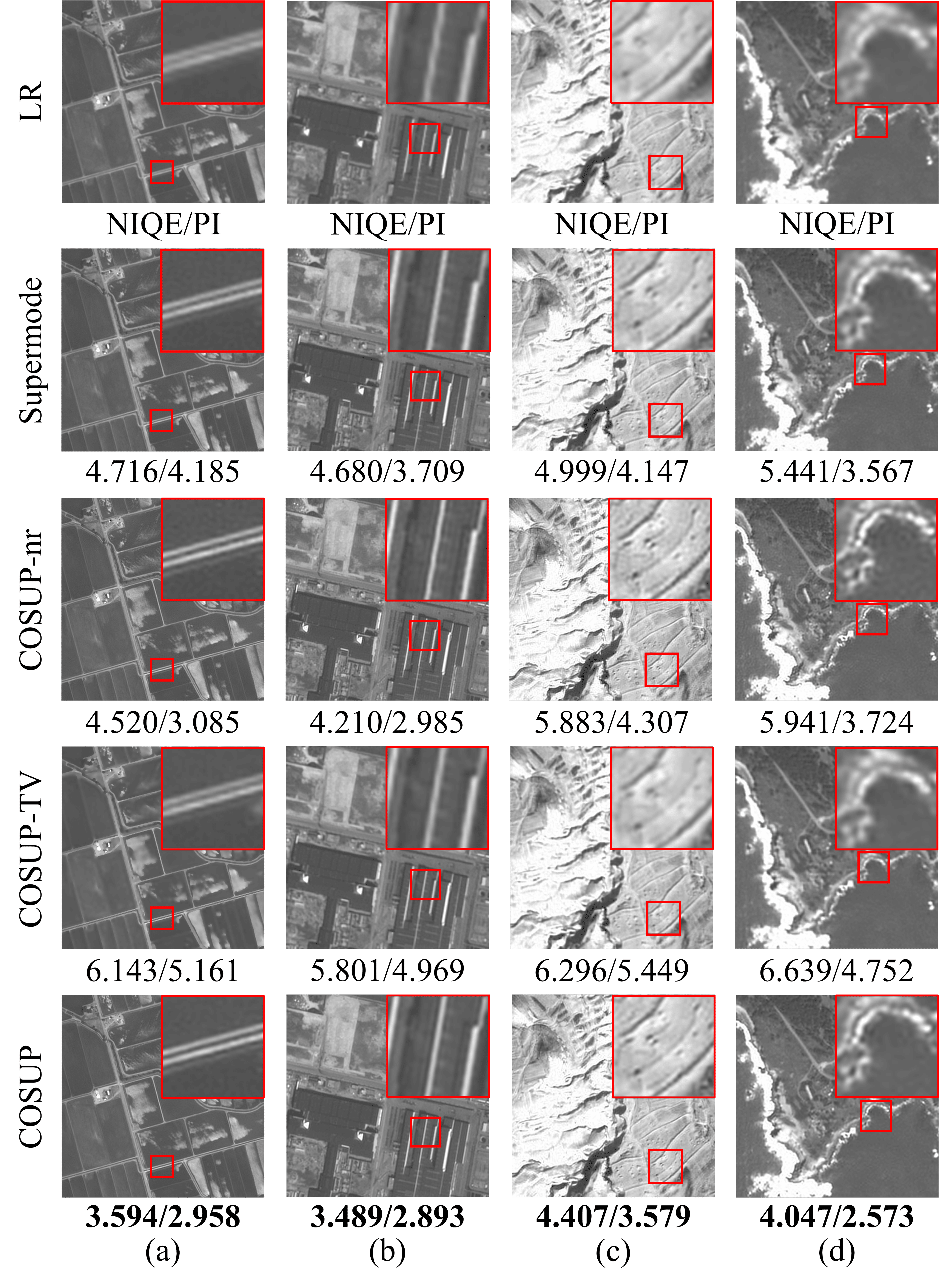}
\caption{Visual results on real SPOT-5 data with corresponding NIQE/PI values for four representative scenes, including (a) farm, (b) city, (c) mountain, and (d) coastline. 
We demonstrate the LR image ($\by_1$), CNES supermode product, as well as the TISR results from COSUP-nr (nr: no regularization), COSUP-TV, and COSUP, respectively.}
\label{realdata_figs}
\end{figure}

\begin{figure}[!t]
\centering
\hspace*{-1.7mm}\includegraphics[width=1.01\linewidth]{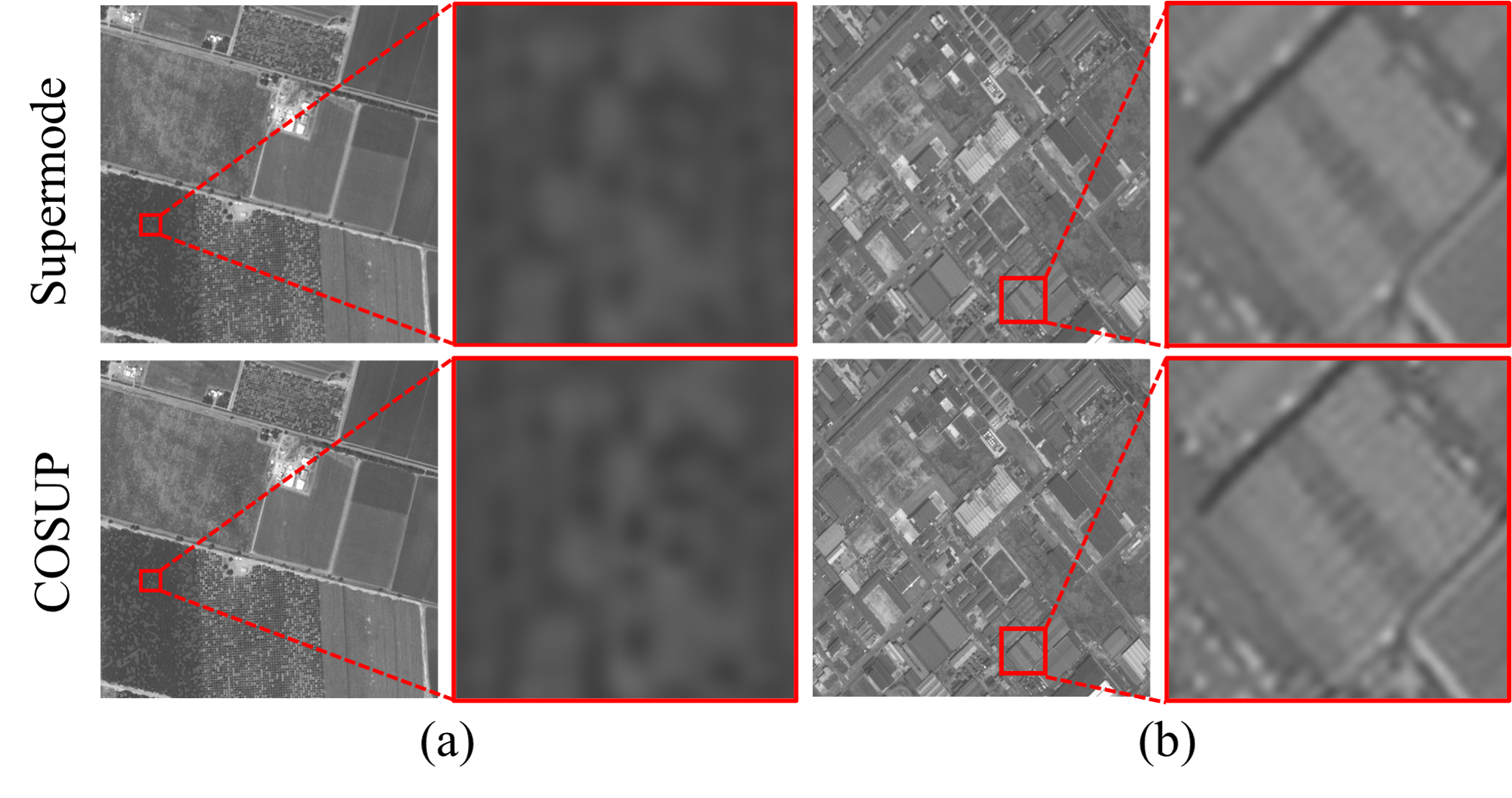}
\caption{ The close-up comparison between the CNES supermode product and the COSUP results on real SPOT-5 data over (a) farm scene and (b) city scene.
}
\label{Fidelity_figs}
\end{figure}

The strong performance of COSUP stems from the adopted convex self-similarity regularization.
To see it, we conduct an ablation study wherein the convex self-similarity module (i.e., the Swin-T module) is removed, and the resulting method is called COSUP-nr (nr: no regularization).
Furthermore, we substitute the Swin-T module with the most popularly adopted TV prior \cite{TV}, denoting this variant as COSUP-TV, to evaluate the reconstruction performance against frequently used priors.
As shown in Table~\ref{tab:no_ref_metrics}, COSUP consistently outperforms the COSUP-nr in terms of NIQE and PI as expected, demonstrating that the incorporation of self-similarity attention effectively enhances reconstruction quality. 
COSUP-nr even occasionally gets weaker performances compared to the official supermode products.
Notably, COSUP-TV exhibits inferior performance across all scenes, further indicating that our convex self-similarity prior does provide stronger structural information that is crucial for the TISR performances.
Overall, the proposed COSUP algorithm generally achieves superior results across different land-cover categories, reflecting its robustness and generalization capability in the challenging real data scenarios (cf. Figure \ref{fig:Shift}).

Figure~\ref{realdata_figs} further presents visual comparisons among the input LR image ($\by_1$), the official supermode product, the COSUP-nr reconstruction, as well as the proposed COSUP result over representative landscapes. 
Overall, COSUP well preserves fine details and sharp boundaries more reliably, echoing the best NIQE and PI values highlighted in boldfaced numbers (cf. Figure~\ref{realdata_figs}). 
In Figure~\ref{realdata_figs}(a), COSUP clearly separates the two slender roads between the fields, and maintains continuous and well-defined lines, whereas the supermode product is relatively less effective in visually distinguishing them.  
In Figure~\ref{realdata_figs}(b), COSUP distinctly reconstructs the fine periodic black stripes between rooftops and clearly preserves the white stripes.
In Figure~\ref{realdata_figs}(c), compared with the official supermode product and COSUP-TV, COSUP demonstrates superior capability in recovering nonlinear textures and sharply delineating their contours.
%
%
In Figure~\ref{realdata_figs}(d), COSUP maintains smooth shoreline contours while preserving interior textures and small highlights.
Although COSUP and COSUP-nr appear broadly similar in overall visual quality, local differences can still be observed in certain regions of the red frames; for example, within the central white stripe region of Figure~\ref{realdata_figs}(b), COSUP-nr shows slightly uneven and inconsistent structures in some edges and textures.
In Figure~\ref{realdata_figs}(c) and Figure~\ref{realdata_figs}(d), COSUP-nr also produces some unnatural dark spots.
As for COSUP-TV, it exhibits noticeable over-smoothing and a loss of edge details across all visual comparisons due to the characteristics of TV, resulting in the weakest performance among all methods.
These artifacts (in COSUP-nr) and over-smoothed textures (in COSUP-TV) introduce statistical inconsistencies, which are reflected in the NIQE and PI measures (cf. Figure \ref{realdata_figs}).
In contrast, COSUP more effectively maintains consistency across the entire image, thereby achieving superior NIQE and PI performances.
%

Although NIQE and PI are widely used as standard no-reference quality metrics \cite{10912673, 10562345, 9861602, 10353979}, they are originally designed based on the statistical characteristics of natural images and may not fully reflect the radiometric and geometric fidelity of remote sensing imagery.
Therefore, we further evaluate the preservation of spatial structures and boundary clarity by conducting detailed close-up inspections of the proposed COSUP and the baseline supermode product, as demonstrated in Figure \ref{Fidelity_figs}.
In the farm scene [cf. Figure \ref{Fidelity_figs}(a)], COSUP more accurately delineates the spacing between tree crops, which appear as distinct dark spots (they are more intermingled together in the supermode product).
In Figure \ref{Fidelity_figs}(b), COSUP preserves building roof boundaries with greater sharpness.
These observations again indicate that the proposed convex self-similarity regularization strategy effectively maintains spatial structural consistency and boundary sharpness, thereby achieving superior geospatial fidelity.
%

These visual observations are consistent with the outstanding quantitative performances of COSUP (cf. Table~\ref{tab:no_ref_metrics}), confirming the robustness of COSUP on real SPOT-5 data even in the presence of nonideal subpixel shifts.
Remarkably, COSUP is trained using data from other satellites (cf. Table \ref{tab:dataset_comparison}), while it still exhibits strong performance on SPOT-5 satellite data, illustrating the outstanding generalization capability.
Finally, the proposed COSUP is computationally very efficient (cf. Table \ref{tab:psnr_ssim_by_scene} and Table \ref{tab:no_ref_metrics}).

%

\section{Conclusion and Future Works}\label{sec:conclusion}

Echoing the recent trend in developing interpretable artificial intelligence, we have developed an interpretable COSUP algorithm for the TISR task, which has critical remote sensing applications such as the supermode imaging.
We formulated the TISR problem using a convex self-similarity regularizer [cf. \eqref{eq:regularized_problem}], and implemented it using Swin-Transformer under the deep unfolding framework.
The more generalized convex self-similarity function (cf. Appendix \ref{appdx:Kalpha}) is applicable not only to TISR, but also to those ill-posed problems that require spatial regularization in the future.
Although closed-form expression of the proximal mapping associated with the convex self-similarity function is derivable, its deep unfolding implementation yields a millisecond-level near-real-time SR computing. 
As it turns out, the proposed COSUP algorithm significantly outperforms existing baselines with outstanding generalization ability.
For experiments on real satellite data with spatially nonideal subpixel shifts, the SR results of COSUP still exhibit great superiority over the official CNES supermode imaging products in terms of credible metrics (i.e., NIQE and PI).

As illustrated in Figure \ref{fig:shift_curve}, the performance of COSUP tends to degrade as subpixel shifts deviate from the ideal offset (i.e., 0.5-pixel shift).
However, even for large subpixel shifts, the proposed COSUP algorithm still preserves sufficiently strong super-resolution performances with PSNR values over 40dB (cf. Figure \ref{fig:shift_curve}).
Given that real-world satellite imaging systems may encounter more severe pixel displacements due to platform vibrations or registration errors, the super-resolution results would be degraded with ghosting artifacts.
Therefore, an important future research line is to extend the COSUP theory to robustly handle scenarios with severe pixel displacements, thereby further broadening the application scenarios.
On the other hand, quantum deep learning is critical for capturing fine details, as recently validated on the change detection and weather forecasting applications \cite{QUEENG,SQUAREMAMBA}, and hence will also be a promising future research line for the target super-resolution tasks.

\appendix

\subsection{A More Generalized Definition of Convex Self-Similarity Pattern for the TISR Problem}\label{appdx:Kalpha}

In this appendix, we explain how to learn the self-similarity pattern $(\mathcal{K},\{\alpha_{i,j}\})$ in the TISR case.
Note that in the TISR problem setting, we do not have the HR reference $\bz$ of the convex function \eqref{eq:ss_phi}.
Fortunately, inspired by Figure \ref{fig:TISRillustration}, we notice that the self-similarity pattern of $\bz$ should be structurally equivalent to that of $\by_1$.
Specifically, if two patches are similar in $\bz$, their corresponding downsampled versions in $\by_1$ must also be similar.
%
However, the converse may not always be true.
%
Due to the loss of high-frequency details, dissimilar HR patches might appear to be similar in the LR domain. 
Although relying solely on $\by_1$ to determine self-similarity inevitably introduces some potential estimation bias, it remains the most viable approximation under the traditional convex optimization setting.
Although we can choose LR $\by_1$ as the reference image (i.e., $\by_{\text{ref}} =\by_1$) in the TISR task, we will propose a more judicious strategy that uses HR self-similarity information under the deep learning setting, as will be discussed in Appendix \ref{appdx:trainSwinT}.
%

In Appendix \ref{appdx:Kalpha}, however, we focus on building the basic definition of convex self-similarity function.
The graph $\mathcal{K}$ then contains all pairs of similar patches as its edges, and $\alpha_{i,j} \geq 0$ denotes the similarity weight of the edge $(i,j)$.
The $k$th node in the graph $\mathcal{K}$ represents the $k$th patch.
To quantify the patch similarity, we define the edge weight $\alpha_{i,j}$ as
\begin{equation}
\alpha_{i,j}
\triangleq
\left\| \mathbf{P}_i \by_{\text{ref}} - \mathbf{P}_j \by_{\text{ref}} \right\|_2^{-1},
\label{eq:alpha}
\end{equation}
where $\mathbf{P}_k$ is the operator to sift the $k$th patch from $\by_{\text{ref}}$.
The meaning of \eqref{eq:alpha} is that as the $i$th and $j$th patches are more similar, the Euclidean norm becomes smaller, implying a larger similarity weight $\alpha_{i,j}$ of edge $(i,j)$.

Therefore, with the similarity score $\alpha_{i,j}$, the graph $\mathcal{K}$ can be constructed by searching $k$ most similar patches for each patch $i$.
Mathematically, we can define the graph $\mathcal{K}$ as
\begin{equation}\label{eq:K}
\mathcal{K} \triangleq \left\{ (i,j) \;\middle|\; i \in \mathcal{I},\ j \in \mathcal{N}_k(i) \right\},
\end{equation}
where $\mathcal{I}$ is the index set with each index $i\in\mathcal{I}$ representing a patch, and the set $\mathcal{N}_k(i)\subseteq\mathcal{I}$ collects the $k$ indices of the $k$ patches that are most similar to the patch $i$.
To be rigorous, we have
\[
\mathcal{N}_k(i)
\in 
\arg\max_{ \mathcal{N}\subseteq\mathcal{I},~|\mathcal{N}|=k }~\sum_{j\in\mathcal{N}}\alpha_{i,j},
\]
where the model order is empirically set as $k:=3$.
We remark that when $k:=1$, the above definition reduces back to the naive one originally proposed in \cite{SSSS}.
Thus, this is a more generalized definition of convex self-similarity prior, and is applicable to the TISR task.

\subsection{Training Details of Swin-T Embedded in the COSUP Unfolding Network}\label{appdx:trainSwinT}

In the pretraining phase of Swin-T, denoted as $G(\cdot)$, its input is an HR image $\widetilde{\bz}$, and the output $G(\widetilde{\bz})$ is the denoised version.
According to \eqref{eq:ss_phi}, the self-similarity loss function w.r.t. the HR image $\widetilde{\bz}$ can then be defined as
\begin{equation*}
\begin{aligned}
\lVert G(\widetilde{\bz}) - \widetilde{\bz} \rVert_1
+ 
\frac{\lambda}{2} 
\sum_{(i,j) \in \mathcal{K}} \widetilde{\alpha}_{i,j} 
\, \big\| \mathcal{P}_i\big(G(\widetilde{\bz})\big) - \mathcal{P}_j\big(G(\widetilde{\bz})\big) \big\|_2^2.
\end{aligned}
\end{equation*}
Here, $ \lambda:=0.1$ is the regularization weight of the self-similarity term; 
$\mathcal{P}_i(\cdot)$ and $\mathcal{P}_j(\cdot)$ denote the network extraction operators for the $i$th and $j$th HR patches, corresponding to the patch-sifting operators $\mathbf{P}_i$ and $\mathbf{P}_j$ in \eqref{eq:ss_phi}, respectively; 
$\mathcal{K}$ denotes the self-similarity network corresponding to the HR training data $\widetilde{\bz}$; and 
$\widetilde{\alpha}_{i,j}$ is the HR self-similarity degree (between the $i$th and $j$th HR patches) computed from HR $\widetilde{\bz}$.
The L1-norm term is to reflect the nature of a denoising network, whose input and output are expected to hold sufficient resemblance, and is known to be more robust against outliners compared to the L2-norm data-fitting strategy.

The above design preserves structural and texture details while accounting for the contribution of local image contrast to self-similarity.
The pretrained Swin-T is then plugged into the overall TISR deep unfolding network deployed in Figure \ref{fig:stage_k}.
This is for guiding the training of the other parts of the COSUP unfolding network (i.e., those parts other than Swin-T), composed of some convolutional layers and residual blocks, etc.
Each ResBlock contains two $3\times3$ convolutional layers with an intermediate ReLU activation, and employs identity skip connections to maintain feature consistency and stability, as illustrated in Figure \ref{fig:swinT}.
The other parts of the deep unfolding network with a model-order of $K:=3$ have nothing to do with the self-similarity proximal mapping, and hence is trained using only the L1-norm loss term.
Since the parameters of the pretrained Swin-T remain frozen when training the other parts of the network, its pretrained self-similarity structural prior is judiciously preserved.

\bibliography{ref}

@article{an2022tr,
  title={{TR-MISR}: Multiimage super-resolution based on feature fusion with transformers},
  author={An, Tai and Zhang, Xin and Huo, Chunlei and Xue, Bin and Wang, Lingfeng and Pan, Chunhong},
  journal={IEEE Journal of Selected Topics in Applied Earth Observations and Remote Sensing},
  volume={15},
  pages={1373--1388},
  month = {Jan.},
  year={2022},
  publisher={IEEE}
}

@article{SQUAREMAMBA,
  title={A Quantum-Empowered {SPEI} Drought Forecasting Algorithm Using Spatially-Aware Mamba Network},
  author={Tang, Po-Wei and Lin, Chia-Hsiang and Huang, Jian-Kai and Huete, Alfredo R},
  journal={IEEE Transactions on Geoscience and Remote Sensing},
  volume={63},
  pages={1--18},
  year={2025},
  publisher={IEEE}
}

@inproceedings{xu2015patch,
  title={Patch group based nonlocal self-similarity prior learning for image denoising},
  author={Xu, Jun and Zhang, Lei and Zuo, Wangmeng and Zhang, David and Feng, Xiangchu},
  booktitle={Proc. {IEEE} International Conference on Computer Vision},
  pages={244--252},
  month= {Dec. 7--13,},
  year={2015},
  address= {Santiago, Chile},
}

@article{paris2018novel,
  title={A novel sharpening approach for superresolving multiresolution optical images},
  author={Paris, Claudia and Bioucas-Dias, Jose and Bruzzone, Lorenzo},
  journal={IEEE Transactions on Geoscience and Remote Sensing},
  volume={57},
  number={3},
  pages={1545--1560},
  month = {Mar.},
  year={2019},
  publisher={IEEE}
}

@inproceedings{burger2012image,
  title={Image denoising: Can plain neural networks compete with {BM3D}?},
  author={Burger, Harold C and Schuler, Christian J and Harmeling, Stefan},
  booktitle={Proc. {IEEE} Conference on Computer Vision and Pattern Recognition},
  pages={2392--2399},
  month = {Jun. 16--21,},
  year={2012},
  address = {Providence, RI, USA},
}

@article{buades2011self,
  title={Self-similarity-based image denoising},
  author={Buades, Antoni and Coll, Bartomeu and Morel, Jean-Michel},
  journal={Communications of the ACM},
  volume={54},
  number={5},
  pages={109--117},
  month = {May},
  year={2011},
  publisher={ACM New York, NY, USA}
}

@inproceedings{SISRlin2022single,
  title={Single hyperspectral image super-resolution using {ADMM-Adam} theory},
  author={Lin, Tzu-Hsuan and Lin, Chia-Hsiang},
  booktitle={Proc. {IEEE} International Geoscience and Remote Sensing Symposium},
  pages={1756--1759},
  month = {Jul. 17–22,},
  year={2022},
  address = {Kuala Lumpur, Malaysia},
}

@inproceedings{hsu2024hyperqueen,
  title={Hyper{QUEEN-MF}: Hyperspectral quantum deep network with multi-scale feature fusion for quantum image super-resolution},
  author={Hsu, Shih-Min and Lin, Tzu-Hsuan and Lin, Chia-Hsiang},
  booktitle={Proc. IEEE Sensor Array and Multichannel Signal Processing Workshop (SAM)},
  pages={1--5},
  month={Jul. 8–11,},
  year={2024},
  address={Corvallis, OR, USA},
}

@article{lei2021hybrid,
  title={Hybrid-scale self-similarity exploitation for remote sensing image super-resolution},
  author={Lei, Sen and Shi, Zhenwei},
  journal={IEEE Transactions on Geoscience and Remote Sensing},
  volume={60},
  pages={1--10},
month = {Apr.},
  year={2022},
  publisher={IEEE}
}

@ARTICLE{10912673,
  author={Yan, Xinyu and Chen, Jiuchen and Xu, Qizhi and Li, Wei},
  journal={IEEE Transactions on Geoscience and Remote Sensing}, 
  title={Mitigating Texture Bias: A Remote Sensing Super-Resolution Method Focusing on High-Frequency Texture Reconstruction}, 
  year={Mar. 2025},
  volume={63},
  pages={1-18}
}

@ARTICLE{10562345,
  author={Hou, Mingyang and Huang, Zhiyong and Yu, Zhi and Yan, Yan and Zhao, Yunlan and Han, Xiao},
  journal={IEEE Transactions on Geoscience and Remote Sensing}, 
  title={{CSwT-SR}: Conv-{S}win Transformer for Blind Remote Sensing Image Super-Resolution With Amplitude-Phase Learning and Structural Detail Alternating Learning}, 
  year={Jun. 2024},
  volume={62},
  pages={1-14}
}

@ARTICLE{9861602,
  author={Liu, Ziyu and Feng, Ruyi and Wang, Lizhe and Han, Wei and Zeng, Tieyong},
  journal={IEEE Transactions on Geoscience and Remote Sensing}, 
  title={Dual Learning-Based Graph Neural Network for Remote Sensing Image Super-Resolution}, 
  year={Aug. 2022},
  volume={60},
  pages={1-14}
}

@ARTICLE{10353979,
  author={Xiao, Yi and Yuan, Qiangqiang and Jiang, Kui and He, Jiang and Jin, Xianyu and Zhang, Liangpei},
  journal={IEEE Transactions on Geoscience and Remote Sensing}, 
  title={{EDiffSR}: An Efficient Diffusion Probabilistic Model for Remote Sensing Image Super-Resolution}, 
  year={2024},
  volume={62},
  pages={1-14}
}

@article{kim2025self,
  title={Self-learning based joint multi image super-resolution and sub-pixel registration},
  author={Kim, Hansol and Lee, Sukho and Kang, Moon Gi},
  journal={Digital Signal Processing},
  volume={156},
  pages={104837},
  month = {Jan.},
  year={2025},
  publisher={Elsevier}
}

@article{wang2023multi,
  title={Multi-frame super-resolution of remote sensing images using attention-based {GAN} models},
  author={Wang, Peijuan and Sertel, Elif},
  journal={Knowledge-Based Systems},
  volume={266},
  pages={110387},
  month = {Apr.},
  year={2023},
  publisher={Elsevier}
}

@article{irani1991improving,
  title={Improving resolution by image registration},
  author = {Michal Irani and Shmuel Peleg},
  journal = {CVGIP: Graphical Models and Image Processing},
  volume={53},
  number={3},
  month = {May},
  pages={231--239},
  year={1991},
  publisher={Elsevier}
}

@article{kawulok2019deep,
  author={Kawulok, Michal and Benecki, Pawel and Piechaczek, Szymon and Hrynczenko, Krzysztof and Kostrzewa, Daniel and Nalepa, Jakub},
  journal={IEEE Geoscience and Remote Sensing Letters}, 
  title={Deep Learning for Multiple-Image Super-Resolution}, 
  volume={17},
  number={6},
  pages={1062--1066},
  month = {Jun.},
  year={2020},
  publisher={IEEE}
}

@article{stark1989high,
  title={High-resolution image recovery from image-plane arrays, using convex projections},
  author={Stark, Henry and Oskoui, Peyma},
  journal={Journal of the Optical Society of America A},
  volume={6},
  number={11},
  pages={1715--1726},
  month = {Nov.},
  year={1989},
  publisher = {Optica Publishing Group},
}

@inproceedings{lin2024synthesis,
  title={Synthesis of high-resolution {FORMOSAT}-8 satellite image using fast convex deep learning algorithm},
  author={Lin, Chia-Hsiang and Young, Si-Sheng and Chang, Li-Yu and Liu, Cynthia SJ},
  booktitle={Proc. {IEEE} International Geoscience and Remote Sensing Symposium},
  pages={7657--7661},
  month = {Jul. 7–12,},
  year={2024},
  address = {Athens, Greece},

}

@inproceedings{latry1998spot5,
  title={{SPOT5 THR} mode},
  author={Latry, Christophe and Rouge, Bernard},
  booktitle={Proc. SPIE Earth Observing Systems III},
  volume={3439},
  pages={480--491},
  year={1998},
  month={Oct.},
  address={San Diego, CA, USA}
}

@article{hardie1997joint,
  title={Joint {MAP} registration and high-resolution image estimation using a sequence of undersampled images},
  author={Hardie, Russell C and Barnard, Kenneth J and Armstrong, Ernest E},
  journal={IEEE Transactions on Image Processing},
  volume={6},
  number={12},
  pages={1621--1633},
  month = {Dec.},
  year={1997},
  publisher={IEEE}
}

@article{molini2019deepsum,
  title={Deep{SUM}: Deep Neural Network for Super-Resolution of Unregistered Multitemporal Images},
  author={Molini, Andrea Bordone and Valsesia, Diego and Fracastoro, Giulia and Magli, Enrico},
  journal={IEEE Transactions on Geoscience and Remote Sensing},
  volume={58},
  number={5},
  pages={3644--3656},
  month = {May},
  year={2020},
  publisher={IEEE}
}

@article{wu2024hybrid,
  title={A hybrid network of {CNN} and transformer for subpixel shifting-based multi-image super-resolution},
  author={Wu, Qiang and Zeng, Hongfei and Zhang, Jin and Li, Weishi and Xia, Haojie},
  journal={Optics and Lasers in Engineering},
  volume={182},
  pages={108458},
  month = {Nov.},
  year={2024},
  publisher={Elsevier}
}

@article{dong2019selection,
  title={Selection-based subpixel-shifted images super-resolution},
  author={Dong, Lili and Jin, Jie and Jiang, Yuhang and Zhang, Meng and Xu, Wenhai},
  journal={IEEE Access},
  volume={7},
  pages={110951--110963},
  month = {Aug.},
  year={2019},
  publisher={IEEE}
}

@article{sun2025super,
  title={Super-Resolution for Remote Sensing Imagery via the Coupling of a Variational Model and Deep Learning},
  author={Sun, Jing and Shen, Huanfeng and Yuan, Qiangqiang and Zhang, Liangpei},
  journal={IEEE Transactions on Geoscience and Remote Sensing},
  month = {Feb.},
  year={2025},
  volume={63},
  pages={1-19},
  publisher={IEEE}
}

@article{HyperQUEEN,
  title={Hyper{QUEEN}: {H}yperspectral quantum deep network for image restoration},
  author={Lin, Chia-Hsiang and Chen, You-Yao},
  journal={IEEE Transactions on Geoscience and Remote Sensing},
  volume={61},
  pages={1--20},
  year={May 2023},
  publisher={IEEE}
}

@article{HyperKING,
  title={Hyper{KING}: {Q}uantum-Classical Generative Adversarial Networks for Hyperspectral Image Restoration},
  author={Lin, Chia-Hsiang and Young, Si-Sheng},
  journal={IEEE Transactions on Geoscience and Remote Sensing},
  volume={63},
  pages={1--19},
  year={Apr. 2025},
  publisher={IEEE}
}

@article{COS2A,
  title={{COS2A}: Conversion from {S}entinel-2 to {AVIRIS} hyperspectral data using interpretable algorithm with spectral-spatial duality},
  journal={IEEE Transactions on Geoscience and Remote Sensing}, 
  author        = {Lin, Chia-Hsiang and Chen, J.-T. and Leng, Z.-C. and Lin, J.-T.},
  month = {Oct.},
  year={2025},
  volume={63},
  pages={1-16}
}

@article{zhang2021plug,
  title={Plug-and-play image restoration with deep denoiser prior},
  author={Zhang, Kai and Li, Yawei and Zuo, Wangmeng and Zhang, Lei and Van-Gool, Luc and Timofte, Radu},
  journal={IEEE Transactions on Pattern Analysis and Machine Intelligence},
  volume={44},
  number={10},
  pages={6360--6376},
  year={Jun. 2021},
  publisher={IEEE}
}

@article{parikh2014proximal,
  title={Proximal algorithms},
  author={Parikh, Neal and Boyd, Stephen},
  journal={Foundations and Trends{\textregistered} in Optimization},
  volume={1},
  number={3},
  pages={127--239},
  year={Jan. 2014},
  publisher={Now Publishers, Inc.}
}

@article{zhang2017beyond,
  title={Beyond a {Gaussian} denoiser: Residual learning of deep {CNN} for image denoising},
  author={Zhang, Kai and Zuo, Wangmeng and Chen, Yunjin and Meng, Deyu and Zhang, Lei},
  journal={IEEE Transactions on Image Processing},
  volume={26},
  number={7},
  pages={3142--3155},
  year={Feb. 2017},
  publisher={IEEE}
}

@article{PRIME,
  title={{PRIME}: Blind multispectral unmixing using virtual quantum prism and convex geometry},
  author={Lin, Chia-Hsiang and Lin, Jhao-Ting},
  journal={IEEE Transactions on Geoscience and Remote Sensing},
  volume={63},
  pages={1--15},
  year={Feb. 2025},
  publisher={IEEE}
}

@inproceedings{kingma2014adam,
  author    = {Kingma, D. P.  and
               Ba, J. },
  title     = {{Adam}: {A} Method for Stochastic Optimization},
  booktitle = {Proc. International Conference on Learning Representations},
  year      = {May 7-9, 2015},
  address={San Diego, CA, USA}
}

@article{SSSS,
  title={An Explicit and Scene-Adapted definition of convex self-similarity prior with application to unsupervised {S}entinel-2 super-resolution},
  author={Lin, Chia-Hsiang and Bioucas-Dias, Jos{\'e} M},
  journal={IEEE Transactions on Geoscience and Remote Sensing},
  volume={58},
  number={5},
  pages={3352--3365},
  year={Dec. 2019},
  publisher={IEEE}
}

@book{CVXbookCLL2016,
  title={Convex Optimization for Signal Processing and Communications: From Fundamentals to Applications},
  author={Chi, Chong-Yung and Li, Wei-Chiang and Lin, Chia-Hsiang},
  year={2017},
  publisher={Boca Raton, FL, USA: CRC Press}
}

@article{CODE,
  title={{ADMM-ADAM}: {A} new inverse imaging framework blending the advantages of convex optimization and deep learning},
  author={Lin, Chia-Hsiang and Lin, Yen-Cheng and Tang, Po-Wei},
  journal={IEEE Transactions on Geoscience and Remote Sensing},
  volume={60},
  pages={1--16},
  year={Sep. 2021},
  publisher={IEEE}
}

@article{AAHCSD,
  title={All-addition hyperspectral compressed Sensing for metasurface-driven miniaturized satellite},
  author={Lin, Chia-Hsiang and Lin, Tzu-Hsuan},
  journal={IEEE Transactions on Geoscience and Remote Sensing},
  volume={60},
  pages={1--15},
  year={Mar. 2021},
  publisher={IEEE}
}

@ARTICLE{COCNMF,
  author={Lin, Chia-Hsiang and Ma, Fei and Chi, Chong-Yung and Hsieh, Chih-Hsiang},
  journal={IEEE Transactions on Geoscience and Remote Sensing}, 
  title={A Convex Optimization-Based Coupled Nonnegative Matrix Factorization Algorithm for Hyperspectral and Multispectral Data Fusion}, 
  year={Nov. 2017},
  volume={56},
  number={3},
  pages={1652-1667},
  doi={10.1109/TGRS.2017.2766080}}

@ARTICLE{QUEENG,
  author={Lin, Chia-Hsiang and Lin, Tzu-Hsuan and Chanussot, Jocelyn},
  journal={IEEE Transactions on Geoscience and Remote Sensing}, 
  title={Quantum Information-empowered Graph Neural Network for Hyperspectral Change Detection}, 
  year={Nov. 2024},
  volume={62},
  number={},
  pages={1--15},
  doi={10.1109/TGRS.2024.3490703}}

@ARTICLE{CODEIF,
  author={Lin, Chia-Hsiang and Hsieh, Cheng-Ying and Lin, Jhao-Ting},
  journal={IEEE Transactions on Geoscience and Remote Sensing}, 
  title={{CODE-IF}: {A} Convex/Deep Image Fusion Algorithm for Efficient Hyperspectral Super-Resolution}, 
  year={Apr. 2024},
  volume={62},
  number={},
  pages={1--18},
  doi={10.1109/TGRS.2024.3384808}}

@article{NCCODE,
  title={Metasurface-empowered snapshot hyperspectral imaging with convex/deep {(CODE)} small-data learning theory},
  author={Lin, Chia-Hsiang and Huang, Shih-Hsiu and Lin, Ting-Hsuan and Wu, Pin-Chieh},
  journal={Nature Communications},
  volume={14},
  number={1},
  pages={1--10},
  year={Nov. 2023},
  publisher={Nature Publishing Group UK London}
}

@article{guizar2008efficient,
  author = {Manuel Guizar-Sicairos and Samuel T. Thurman and James R. Fienup},
  journal = {Optics Letters},
  number = {2},
  pages = {156--158},
  publisher = {Optica Publishing Group},
  title = {Efficient subpixel image registration algorithms},
  volume = {33},
  month = {Jan.},
  year = {2008},
  doi = {10.1364/OL.33.000156},
}

@misc{pandata,
  title={{OpenAerialMap}},
  url={https://openaerialmap.org/}
}

@misc{spot5data,
  title={{SPOT} {World} {Heritage}},
  url={https://regards.cnes.fr/user/swh/modules/60/}
}

@misc{swh2acarto,
  title={{SWH}-{2A}-{Carto}},
  url={https://swh-2a-carto.fr/}
}

@article{mittal2013niqe,
  author    = {Mittal, Anish and Soundararajan, Rajiv and Bovik, Alan C.},
  title     = {Making a ``Completely Blind'' Image Quality Analyzer},
  journal   = {IEEE Signal Processing Letters},
  volume    = {20},
  number    = {3},
  pages     = {209--212},
  month = {Mar.},
  year      = {2013},
  doi       = {10.1109/LSP.2012.2227726}
}

@InProceedings{blau2018pirm,
  author = {Blau, Yochai and Mechrez, Roey and Timofte, Radu and Michaeli, Tomer and Zelnik-Manor, Lihi},
  title = {The 2018 {PIRM} Challenge on Perceptual Image Super-Resolution},
  booktitle = {Proc. European Conference on Computer Vision Workshops},
  month= {Sep. 8--14,},
  year = {2018},
  address   = {Munich, Germany}
}

@INPROCEEDINGS{wang2024man,
  author={Wang, Yan and Li, Yusen and Wang, Gang and Liu, Xiaoguang},
  booktitle={Proc. IEEE/CVF Conference on Computer Vision and Pattern Recognition Workshops}, 
  title={Multi-scale Attention Network for Single Image Super-Resolution}, 
  month={Jun. 17--18,},
  year={2024},
  address={Seattle, WA, USA},
  pages={5950--5960},
  doi={10.1109/CVPRW63382.2024.00602}
}

@ARTICLE{farsiu2004fast,
  author={Farsiu, S. and Robinson, M.D. and Elad, M. and Milanfar, P.},
  journal={IEEE Transactions on Image Processing}, 
  title={Fast and robust multiframe super resolution},
  month = {Oct.},
  year={2004},
  volume={13},
  number={10},
  pages={1327--1344},
  doi={10.1109/TIP.2004.834669}
}

@INPROCEEDINGS{liu2021swin,
  author={Liu, Ze and Lin, Yutong and Cao, Yue and Hu, Han and Wei, Yixuan and Zhang, Zheng and Lin, Stephen and Guo, Baining},
  booktitle={Proc. IEEE/CVF International Conference on Computer Vision}, 
  title={Swin {T}ransformer: Hierarchical Vision Transformer using Shifted Windows}, 
  month={Oct. 11--17,},
  year={2021},
  address = {Montreal, QC, Canada},
  pages={9992--10002},
  doi={10.1109/ICCV48922.2021.00986}
}

@article{chang2015pi,
  author = {H. Wen Chang and Q. Wen Zhang and Q. Gang Wu and Y. Gan}, 
  title = {Perceptual image quality assessment by independent feature detector},
  journal = {Neurocomputing},
  volume = {151},
  pages = {1142--1152},
  month = {Mar.},
  year = {2015},
  issn = {0925-2312},
  doi = {https://doi.org/10.1016/j.neucom.2014.04.081},
}

@book{poynton2012digital,
  author    = {Charles Poynton},
  title     = {Digital Video and HD: Algorithms and Interfaces},
  edition   = {2nd},
  publisher = {Morgan Kaufmann},
  year      = {2012},
  address   = {Boston, MA},
  isbn      = {9780123919267},
  url       = {https://www.sciencedirect.com/book/9780123919267/digital-video-and-hd},
}

@misc{formosat-8,
  title={{TASA}: {FORMOSAT}-8},
  url={https://www.tasa.org.tw/zh-TW/missions/detail/FORMOSAT-8}
}

@article{cao2024unsupervised,
  title={Unsupervised hybrid network of transformer and {CNN} for blind hyperspectral and multispectral image fusion},
  author={Cao, Xuheng and Lian, Yusheng and Wang, Kaixuan and Ma, Chao and Xu, Xianqing},
  journal={IEEE Transactions on Geoscience and Remote Sensing},
  volume={62},
  pages={1--15},
  year={2024},
  month={Jan.}
}

@article{CAO2026112374,
  title = {Cross-domain-aware deep unfolding transformer for hyperspectral image super-resolution},
  journal = {Pattern Recognition},
  volume = {172},
  pages = {112374},
  month = {Apr.},
  year = {2026},
  author = {Xuheng Cao and Xuquan Wang and Xiong Dun and Yusheng Lian and Xinbin Cheng and Xiaopeng Hao},
}

@ARTICLE{10764782,
  author={Yao, Xudong and Zhang, Haopeng and Wen, Sizhe and Shi, Zhenwei and Jiang, Zhiguo},
  journal={IEEE Journal of Selected Topics in Applied Earth Observations and Remote Sensing}, 
  title={Single-Image Superresolution for {RGB} Remote Sensing Imagery via Multiscale {CNN-T}ransformer Feature Fusion}, 
  month = {Nov.},
  year={2025},
  volume={18},
  pages={1302--1316}
}

@Article{rs15040882,
AUTHOR = {Zhang, Haopeng and Zhang, Cong and Xie, Fengying and Jiang, Zhiguo},
TITLE = {A Closed-Loop Network for Single Infrared Remote Sensing Image Super-Resolution in Real World},
JOURNAL = {Remote Sensing},
VOLUME = {15},
month = {Feb.},
YEAR = {2023},
NUMBER = {4},
pages = {882},

}

@ARTICLE{9151194,
  author={Zhang, Haopeng and Wang, Pengrui and Jiang, Zhiguo},
  journal={IEEE Transactions on Geoscience and Remote Sensing}, 
  title={Nonpairwise-Trained Cycle Convolutional Neural Network for Single Remote Sensing Image Super-Resolution}, 
  month = {May},
  year={2021},
  volume={59},
  number={5},
  pages={4250--4261}
}

@ARTICLE{9656645,
  author={Ma, Qing and Jiang, Junjun and Liu, Xianming and Ma, Jiayi},
  journal={IEEE Transactions on Computational Imaging}, 
  title={Deep Unfolding Network for Spatiospectral Image Super-Resolution}, 
  month = {Dec.},
  year={2022},
  volume={8},
  number={},
  pages={28--40}
}

@ARTICLE{9904907,
  author={Wang, Jiaming and Shao, Zhenfeng and Huang, Xiao and Lu, Tao and Zhang, Ruiqian},
  journal={IEEE Transactions on Computational Imaging}, 
  title={A Deep Unfolding Method for Satellite Super Resolution}, 
  month = {Sep.},
  year={2022},
  volume={8},
  number={},
  pages={933--944}
}

@Article{rs6087491,
AUTHOR = {Shen, Huanfeng and Zhao, Wennan and Yuan, Qiangqiang and Zhang, Liangpei},
TITLE = {Blind Restoration of Remote Sensing Images by a Combination of Automatic Knife-Edge Detection and Alternating Minimization},
JOURNAL = {Remote Sensing},
VOLUME = {6},
month = {Aug.},
YEAR = {2014},
NUMBER = {8},
PAGES = {7491--7521}
}

@article{TV,
  title={Nonlinear total variation based noise removal algorithms},
  author={Rudin, Leonid I and Osher, Stanley and Fatemi, Emad},
  journal={Physica D: Nonlinear Phenomena},
  volume={60},
  number={1–4},
  pages={259--268},
  month={Nov.},
  year={1992}
}

@ARTICLE{217219,
  author={Savakis, A.E. and Trussell, H.J.},
  journal={IEEE Transactions on Image Processing}, 
  title={Blur Identification by Residual Spectral Matching}, 
  month = {Apr.},
  year={1993},
  volume={2},
  number={2},
  pages={141-151},
  doi={10.1109/83.217219}
}

\begin{IEEEbiography}[{\resizebox{0.9in}{!}{\includegraphics[width=1in,height=1.25in,clip,keepaspectratio]{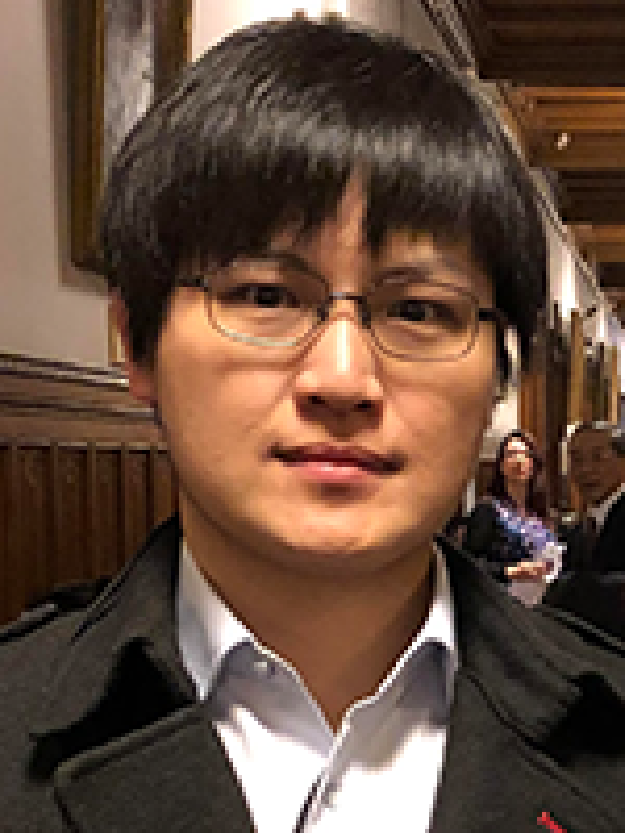}}}]
{\bf Chia-Hsiang Lin}
(S'10-M'18-SM'24)
received the B.S. degree in electrical engineering and the Ph.D. degree in communications engineering from National Tsing Hua University (NTHU), Taiwan, in 2010 and 2016, respectively.
From 2015 to 2016, he was a Visiting Student of Virginia Tech,
Arlington, VA, USA.

He is currently a Professor with the Department of Electrical Engineering,
National Cheng Kung University (NCKU), Taiwan.
Before joining NCKU, he held research positions with The Chinese University of Hong Kong, HK (2014 and 2017),
NTHU (2016-2017),
and the University of Lisbon (ULisboa), Lisbon, Portugal (2017-2018).
He was an Assistant Professor with the Center for Space and Remote Sensing Research, National Central University, Taiwan, in 2018, and a Visiting Professor with ULisboa, in 2019.
His research interests include network science,
quantum computing,
convex geometry and optimization, blind signal processing, and imaging science.

Dr. Lin received the Emerging Young Scholar Award (The 2030 Cross-Generation Program) from National Science and Technology Council (NSTC), from 2023 to 2027,
the Future Technology Award from NSTC, in 2022,
the Outstanding Youth Electrical Engineer Award from The Chinese Institute of Electrical Engineering (CIEE), in 2022,
the Best Young Professional Member Award from IEEE Tainan Section, in 2021,
the Prize Paper Award from IEEE Geoscience and Remote Sensing Society (GRS-S), in 2020,
the Top Performance Award from Social Media Prediction Challenge at ACM Multimedia, in 2020,
and The 3rd Place from AIM Real World Super-Resolution Challenge at IEEE International Conference on Computer Vision (ICCV), in 2019.
He received the Ministry of Science and Technology (MOST) Young Scholar Fellowship, together with the EINSTEIN Grant Award, from 2018 to 2023.
In 2016, he was a recipient of the Outstanding Doctoral Dissertation Award from the Chinese Image Processing and Pattern Recognition Society and the Best Doctoral Dissertation Award from the IEEE GRS-S.
\end{IEEEbiography}

\vspace{-1.3cm}
\begin{IEEEbiography}[{\includegraphics[width=1in,height=1.25in,clip,keepaspectratio]{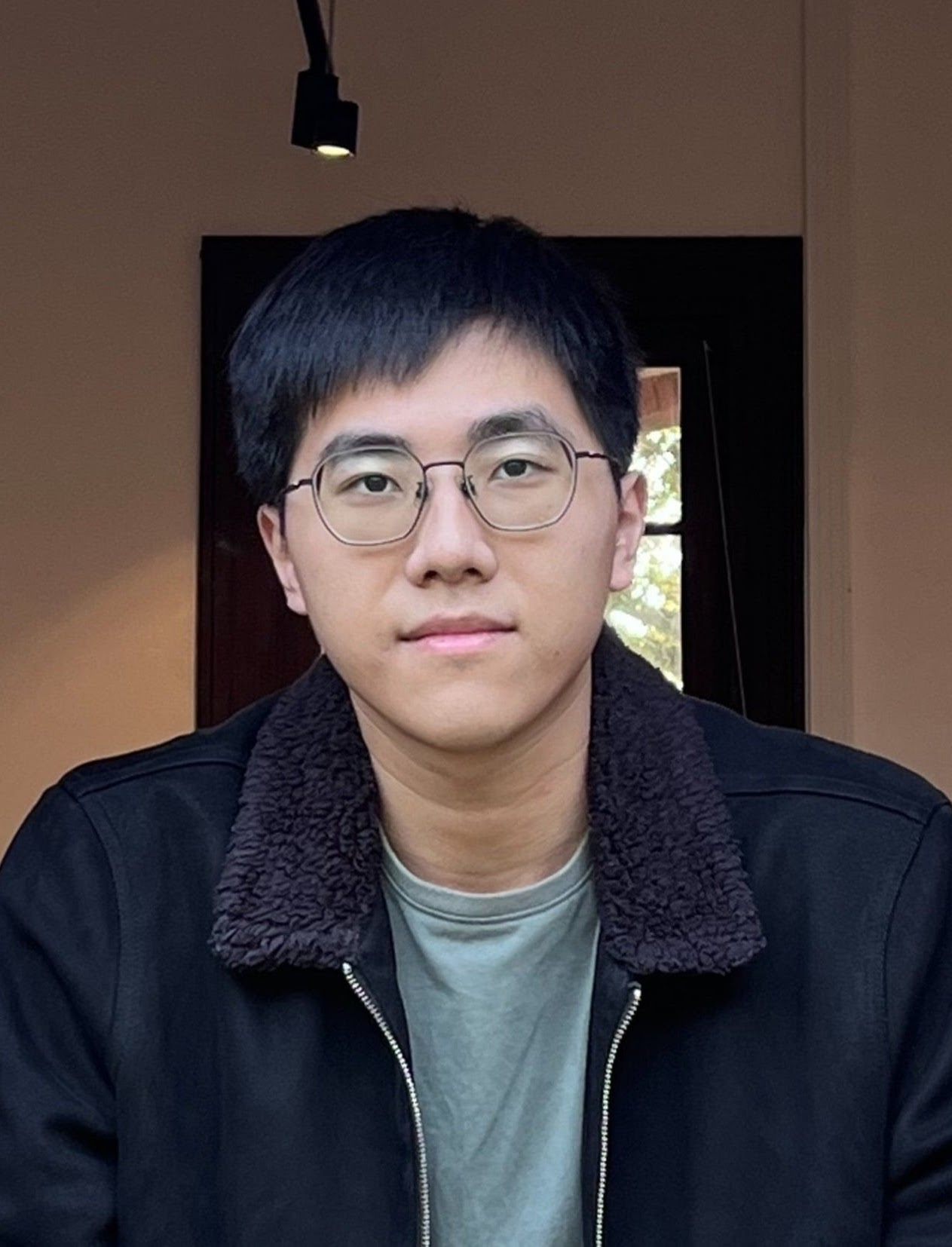}}]{Wei-Chih Liu}
received the B.S. degree from the Department of Civil Engineering, National Cheng Kung University, Tainan, Taiwan, in 2024.

He is currently pursuing the M.S. degree with the Intelligent Hyperspectral Computing Laboratory, Institute of Computer and Communication Engineering, Department of Electrical Engineering, National Cheng Kung University, Tainan, Taiwan. His research interests include convex optimization, deep learning, hyperspectral imaging, and computer vision (e.g., super-resolution and robot vision).
\end{IEEEbiography}
\vspace{-1.3cm}
\begin{IEEEbiography}[{\includegraphics[width=1in,height=1.25in,clip,keepaspectratio]{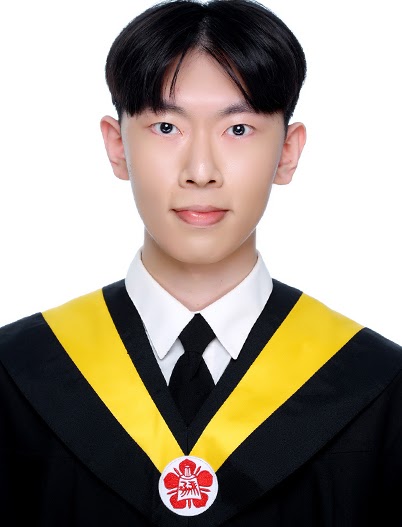}}]{Yu-En Chiu}
received the B.S. degree from the Department of Photonics, National Cheng Kung University, Tainan, Taiwan, in 2024.

He is currently pursuing the M.S. degree with the Intelligent Hyperspectral Computing Laboratory, Institute of Computer and Communication Engineering, Department of Electrical Engineering, National Cheng Kung University, Tainan, Taiwan. His research interests include convex optimization, deep learning, optical imaging, and computer vision (e.g., super-resolution and robot vision).
\end{IEEEbiography}

\vspace{-1.3cm}
\begin{IEEEbiography}[{\resizebox{1in}{!}{\includegraphics[width=1in,height=1.25in,clip,keepaspectratio]{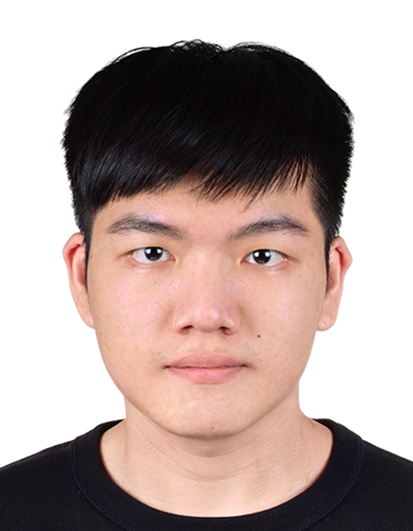}}}]
    	{\bf Jhao-Ting Lin}
		(S'20)
received his B.S. degree from the Department of Communications, Navigation and Control Engineering, National Taiwan Ocean University, Taiwan, in 2020.

He is currently a Ph.D. student affiliated with the Intelligent Hyperspectral Computing Laboratory, Institute of Computer and Communication Engineering, Department of Electrical Engineering, National Cheng Kung University, Taiwan. 
His research interests include convex optimization, deep learning, signal processing, quantum computing, and hyperspectral/multispectral remote sensing.

He has received some highly competitive student awards, including the 2022, 2024, and 2025 Pan Wen Yuan Award from the Industrial Technology Research Institute (ITRI), Taiwan.
He has been selected as a recipient for the Ph.D. Students Study Abroad Program from the National Science and Technology Council (NSTC), Taiwan, for visiting the Okinawa Institute of Science and Technology in 2025-2026 term.
\end{IEEEbiography}

\end{document}